\documentclass{aa}  

\usepackage{graphicx}
\usepackage{txfonts}
\usepackage[T1]{fontenc}
\usepackage{graphicx}	
\usepackage{amsmath}	
\usepackage{multirow}
\usepackage{caption}
\usepackage{subcaption}
\usepackage{float}
\usepackage[colorlinks=true,linkcolor=blue,citecolor=blue]{hyperref}%
\usepackage{threeparttable}
\usepackage{booktabs} 
\begin{document}

   \title{Exploring galactic properties with machine learning}

   \subtitle{Predicting star formation, stellar mass, and metallicity from photometric data}

   \author{F. Z. Zeraatgari
          \inst{1}
          \and
          F. Hafezianzadeh\inst{2}
 \and
Y.-X. Zhang\inst{3}
\and
A. Mosallanezhad\inst{1}
\and
J.-Y. Zhang\inst{3}
          }

   \institute{School of Mathematics and Statistics, Xi'an Jiaotong University, Xi'an, Shaanxi 710049, PR China\\
              \email{fzeraatgari@xjtu.edu.cn}; \email{mosallanezhad@xjtu.edu.cn}
         \and
             Department of physics, Institute for Advanced Studies in Basic Sciences, Zanjan, 45195-1159, Iran
         \and
            CAS Key Laboratory of Optical Astronomy, National Astronomical Observatories, Beijing, 100101, China\\
             \email{E-mail: zyx@bao.ac.cn}
             }

   \date{}

 
  \abstract
   {}
   {We explore machine learning techniques to forecast star formation rate, stellar mass, and metallicity across galaxies with redshifts ranging from 0.01 to 0.3.}
   { Leveraging CatBoost and deep learning architectures, we utilize multiband optical and infrared photometric data from SDSS and AllWISE, trained on the SDSS MPA-JHU DR8 catalogue.}
   {Our study demonstrates the potential of machine learning in accurately predicting galaxy properties solely from photometric data. 
   We achieve minimised root mean square errors, specifically employing the CatBoost model. For star formation rate prediction, we attain a value of $\rm{RMSE_{SFR}} = 0.336$~dex, while for stellar mass prediction, the error is reduced to $\rm{RMSE_{SM}} = 0.206$~dex. Additionally, our model yields a metallicity prediction of $\rm{RMSE_{metallicity}} = 0.097$~dex.
   }
   {These findings underscore the significance of automated methodologies in efficiently estimating critical galaxy properties, amid the exponential growth of multi-wavelength astronomy data. Future research may focus on refining machine learning models and expanding datasets for even more accurate predictions.}


   \maketitle
%
\section{Introduction}
The next generation of extensive multi-wavelength photometric sky surveys, exemplified by missions like 
the Vera C. Rubin Observatory Legacy Survey of Space and Time (LSST; \citealt{Ivezic2019}),
the Euclid Space Telescope (\citealt{LaurIJs2011}), and the China Space Station Telescope (CSST; \citealt{Zhan2011}), are poised to deliver vast datasets containing critical galaxy properties. These observed properties are not only essential metrics but also provide extremely valuable information about various aspects of galaxy evolution, revealing intriguing correlations among them \citep{Brinchmann2004, Tremonti2004, Baldry2008, LaraLopez2010, Mannucci2010, Kravtsov2018}.

The determination of galaxy properties, such as stellar mass, star formation rate, and metallicity, is a complex process that relies on a variety of observational and analytical methods, including the versatile techniques of spectral energy distribution (SED) fitting \citep{Walcher2011, Conroy2013}. Astronomers use a combination of techniques to uncover the mysteries of galaxies. They utilize telescopes and instruments sensitive to a wide range of wavelengths, spanning from $\gamma$ ray, X-ray, ultraviolet, optical bands to infrared and radio bands. These instruments capture diverse emissions from galaxies, revealing the radiance of recently born stars in ultraviolet and optical bands, the infrared signals emitted from dust-covered stellar nurseries, and the faint radiation from old, evolved stars within these cosmic structures. Notably, the near-infrared band, with its longer wavelengths, effectively traces the stellar mass linked to the old population, shedding light on the evolution of galaxies \citep{Kennicutt2012}.

SED fitting involves assumptions about the past star formation history of a galaxy, 
derived from the observed SED and encapsulating a range of astrophysical processes 
\citep{Ciesla2017, Bisigello2016, Bisigello2017}. Utilizing advanced physical models, 
synthetic SEDs closely resembling the observed data allow for the extraction of 
critical galaxy properties, including star formation rate (SFR), stellar mass ($\rm M_{\star}$/SM), 
age, metallicity, and the star formation history. However, the reliance 
on pre-existing data, often in the form of template SEDs, introduces inherent biases, 
influencing the accuracy and interpretation of extracted properties. These biases need 
careful consideration to ensure the fidelity of derived galaxy characteristics 
\citep{Conroy2013, Mitchell2013, Mobasher2015, Laigle2019}.

Computational limitations present a challenge, particularly in handling the numerous variables necessary for generating individual spectra. 
As data projects increase in scale, traditional methods, such as Markov Chain Monte Carlo (MCMC) or forward modelling, demand substantial computational resources for assessing properties across a vast number of observed galaxies \citep{Pacifici2015, Smith2015, Speagle2016, Jones2016}.
This demands more efficient and accurate data analysis, prompting the necessity for innovative approaches such as machine learning-based statistical techniques.

The advancements in exploring galaxy properties have evolved significantly 
alongside the rapid progress of machine learning (ML) and deep learning (DL) techniques 
in recent years, with demonstrated effectiveness in managing extensive datasets 
and revealing novel insights \citep{Ball2010, Allen2019}. Machine learning methods 
not only offer improved computational speed but also the ability to solve complex 
problems. This surge of interest in ML extends across diverse fields 
in astronomy, impacting various research domains and methodologies, such as
star-galaxy classification 
\citep{Philip2002, Ball2006, Vasconcellos2011, Abraham2012, Soumagnac2015, Kim2017, Clarke2020, Nakoneczny2021, Cunha2022, Zeraatgari2023}, 
galaxy morphology classification \citep{Dieleman2015, Abraham2018, Dom2018, Barchi2020, Walmsley2020, Nair2022}, gravitational waves identification \citep{George2018}, gravitational lensing identification \citep{Cheng2020}, estimating photometric redshifts and other physical properties of galaxies \citep{Tagliaferri2003, Masters2015, Krakowski2016, Siudek2018, Hoyle2016, Stensbo-Smidt2017, DIsanto2018, Turner2019, Masters2019, Salvato2019, Bonjean2019, DelliVeneri2019, Surana2020, Mucesh2021, Razim2021, Li2023} 

Before exploring how ML and DL impact our research, it is crucial to have a clear understanding of the fundamental distinctions between them.
Machine learning typically relies on traditional algorithms and statistical models to learn patterns from data. In traditional ML, the feature engineering process is vital, where domain experts manually select or engineer relevant features for the model. 
These models are effective for structured data and situations where meaningful insights can be derived from feature engineering. They perform well with smaller datasets and provide interpretability.
This makes them suitable for scenarios where understanding the decision-making process of the model is essential \citep{Bishop2006}.

On the other hand, DL is a subset of machine learning specifically focussing on neural networks with multiple layers. DL models automatically learn hierarchical features from raw data, eliminating the need for extensive feature engineering. They excel at handling large volumes of unstructured data, like images, and require substantial labeled data for training, often achieving state-of-the-art performance. However, they are considered `black-box' models, posing challenges in explaining how they arrive at specific conclusions. 
Understanding the distinctive characteristics of ML and DL is crucial to maximizing their potential in exploring and comprehending various aspects of galaxies in our research. The fundamental disparity lies in their approach to feature engineering—manual for one and autonomous hierarchical extraction for the other—shaping their distinct roles in influencing the application to comprehend galaxy properties \citep{Goodfellow2016}.

Gaining a comprehensive understanding of galaxy characteristics relies heavily on acquiring optical spectroscopic data, which is often challenging and time-consuming to obtain. 
Previous studies have acknowledged the complexities involved in accurately determining galaxy properties, especially in establishing the correlation between optical and infrared (IR) data and aggregated photometric properties across galaxies \citep{Fogarty2017, Pearson2018}.
In our approach, accurately predicting galaxy properties necessitates 
seamless integration of both optical and IR data. Specifically, 
IR data act as a crucial component in estimating properties, 
serving as a surrogate for stellar mass and portraying the SFR \citep{Leger1984}. 
Optical data play a vital role in providing a comprehensive understanding of these properties, as optical spectra enable precise estimations of SFR and corrections for dust attenuation. Additionally, it offers insights into stellar population ages, metallicities, and mass-to-light ratios, while utilizing colour-magnitude diagrams is instrumental in determining the total stellar mass and estimating metallicity through key indicators \citep{Kennicutt1998, Kennicutt2012}.

Furthermore, our research addresses the challenges of estimating SFR, SM, and metallicity, drawing inspiration from previous studies \citep{Stensbo-Smidt2017, Bonjean2019, DelliVeneri2019, Surana2020}. 
As these preceding studies indicate that relying solely on optical data might be insufficient to accurately constrain these galactic properties \citep{Brescia2014, Stensbo-Smidt2017, DelliVeneri2019}, our approach, therefore, incorporates both optical and IR data to improve the accuracy and reliability of our results.
Additionally, within the current literature, uncertainties emerge due to the absence of a distinct superiority among various ML models, emphasizing the inadequacy of reliable benchmarks within the field.

In this investigation, we aim to evaluate the effectiveness of supervised ML, specifically the CatBoost model, and a deep learning methodology employing the Wide and Deep Neural Network (WDNN) architecture. 
Our approach incorporates optical and IR photometric data to focus on predicting essential galaxy properties—SFR, SM, and metallicity.

Our key objective is to assess the performance and accuracy of these methods in addressing the inherent complexities of estimating these specific galaxy properties. To achieve this, we utilize a combination of magnitudes and colours obtained by cross-matching the SDSS and AllWISE catalogues.
In addition, we explore the impact of observational uncertainties by incorporating observational errors in our analysis. 
Harnessing advanced ML methods, the study focuses on enhancing the accuracy and efficiency of predicting galaxy properties.

This paper is organised as follows. Section \ref{data} provides a detailed description of the data utilized in the study. In Section \ref{method}, the methodology adopted for analysis and evaluation is outlined. Section \ref{model} covers the ML model and the subsequent results derived from these methodologies, particularly focussing on the performance of ML models for the entire galaxies and various galaxy types, along with a comparison of our results with other studies. 
Section \ref{summary} summarises our key findings and their implications.

\begin{figure*}
\centering
        \begin{subfigure}[b]{0.33\textwidth}
                \centering
                \includegraphics[width=\linewidth]{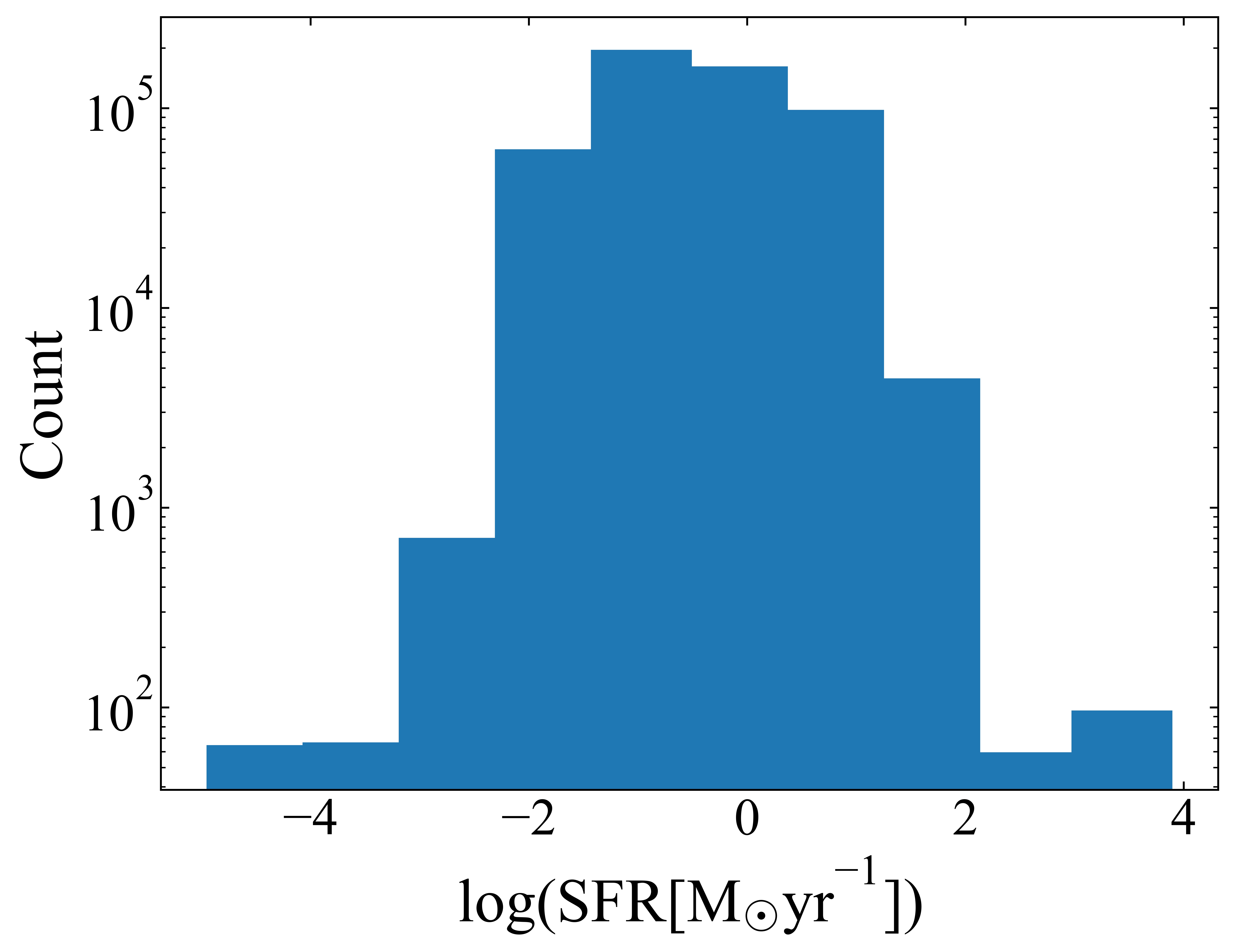}
                \label{fig:QSO}
        \end{subfigure}\hfill
        \begin{subfigure}[b]{0.33\textwidth}
                \centering
                \includegraphics[width=\linewidth]{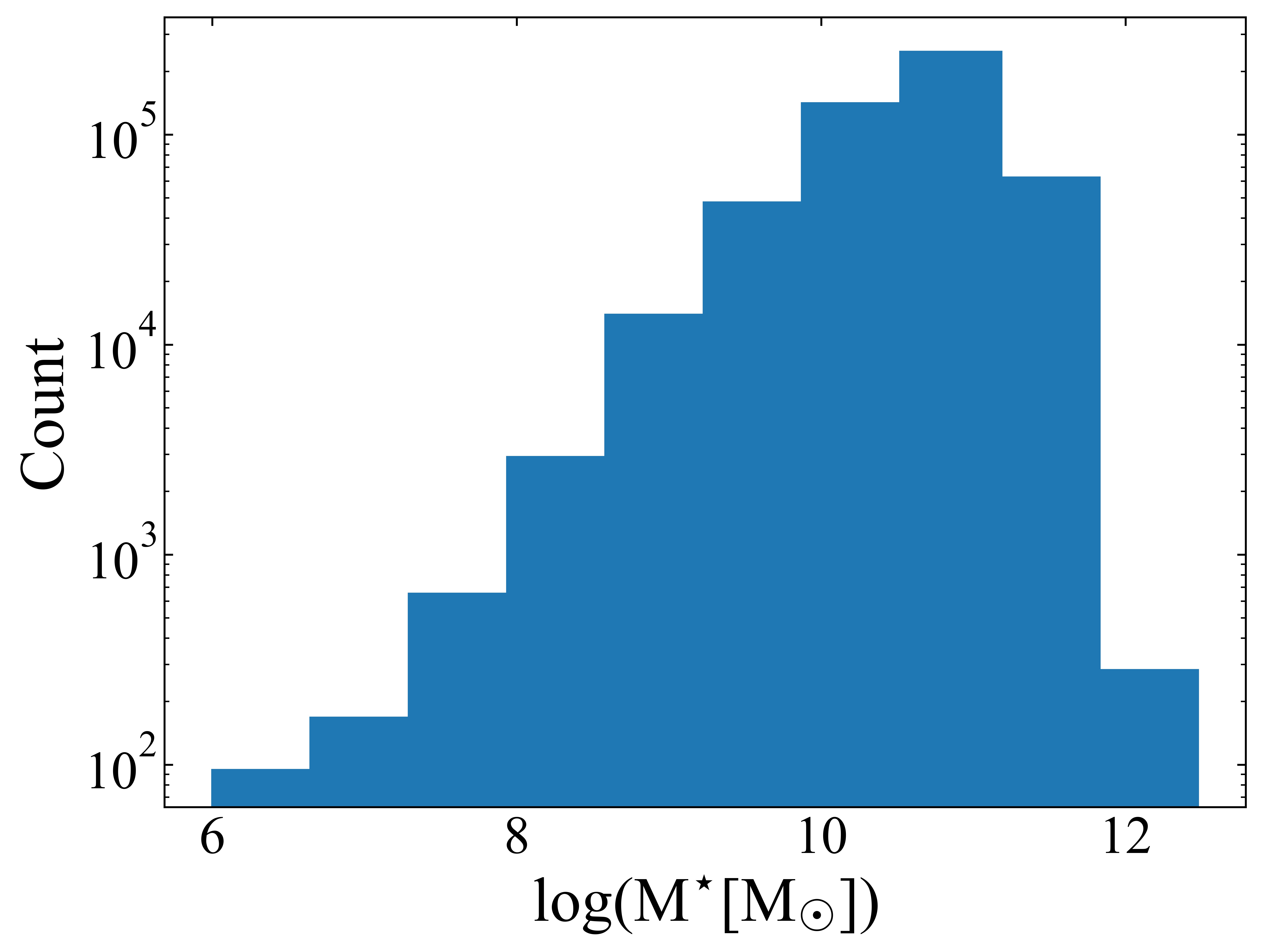}
                \label{fig:NG}
        \end{subfigure}\hfill
        \begin{subfigure}[b]{0.33\textwidth}
                \centering
                \includegraphics[width=\linewidth]{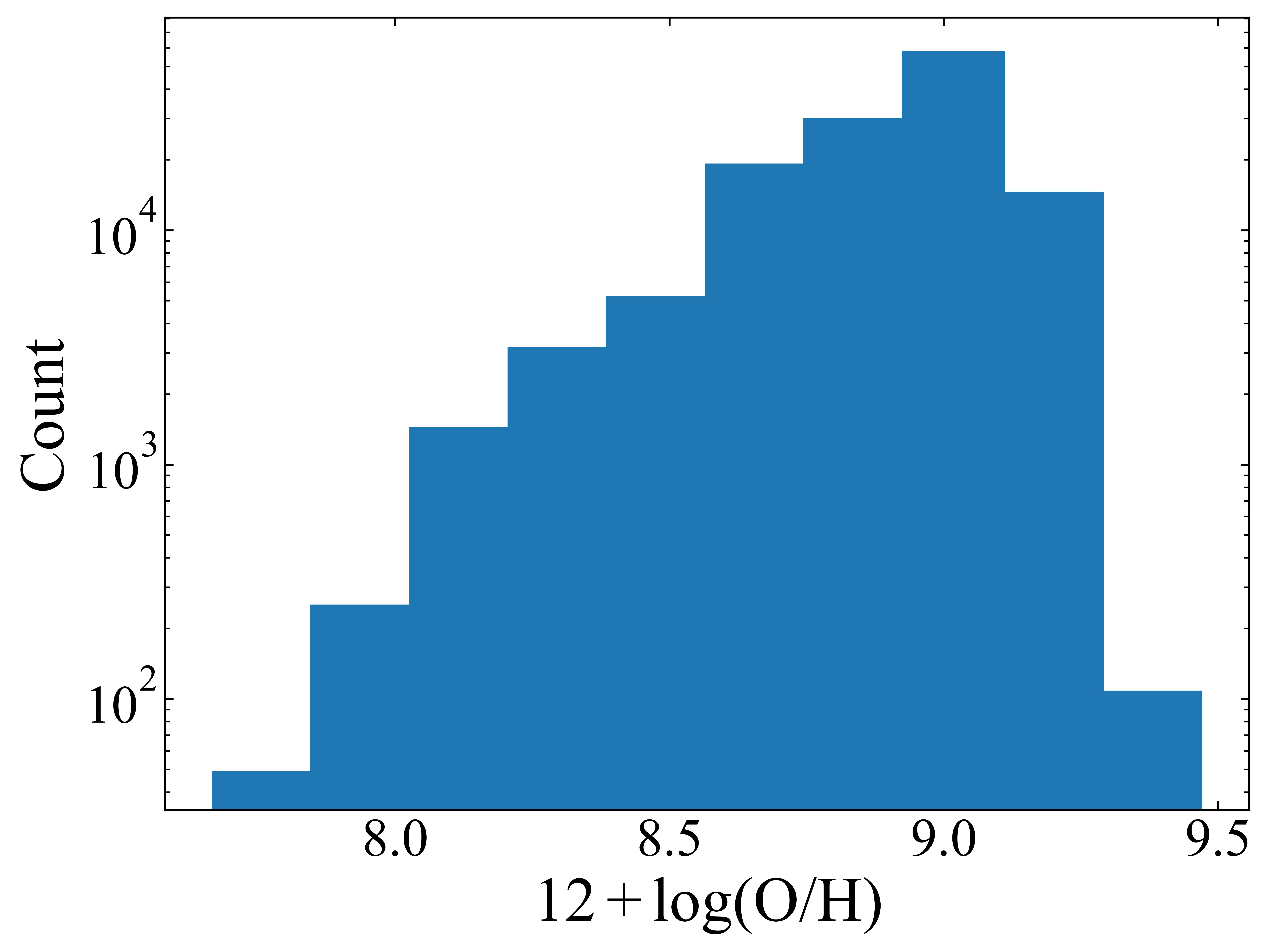}
                \label{fig:EG}
        \end{subfigure}\hfill
        \caption{The distribution of star formation rate (SFR) (left panel), stellar mass (SM) (middle panel), and metallicity (right panel).}\label{fig:distribution}
\end{figure*}
\section{Data}\label{data}
The Sloan Digital Sky Survey (SDSS; \citealt{York2000}) represents one of the most extensive optical surveys in existence, capturing deep images across approximately one-third of the sky in five distinct optical bands: $u$ ($\lambda = 0.355$ $\mu$m), $g$ ($\lambda = 0.477$ $\mu$m), $r$ ($\lambda = 0.623$ $\mu$m), $i$ ($\lambda = 0.762$ $\mu$m), and $z$ ($\lambda = 0.913$ $\mu$m).
In this study, we use the SQL query tool\footnote{https://skyserver.sdss.org/casjobs/} for SDSS to extract optical data from SDSS Data Release 18 (SDSS-DR18).

Our primary focus is on the MPA–JHU DR8 catalogue, a collaborative project between the Max Planck Institute for Astrophysics and Johns Hopkins University \citep{Kauffmann2003, Brinchmann2004}.
This catalogue offers data on SFR, SM, and metallicities for a vast collection of 1~843~200 galaxies, covering a range of redshifts up to approximately $ z \sim 0.3 $. 

The SFR values (flagged as \texttt{SFR\_TOT\_P50}) are determined through the analysis of $ \mathrm{H{\alpha}} $ emission lines whenever they are present. These values are then adjusted to account for dust extinction using the Balmer decrement $ \mathrm{H{\alpha}/H{\beta}} $ as detailed in the work by \cite{Brinchmann2004}. In the absence of emission lines in certain galaxies, SFRs are instead estimated using a relationship between SFR and the spectral index $ \mathrm{D_{4000}} $ as outlined in the studies by \cite{Bruzual1983}, \cite{Balogh1999}, and \cite{Brinchmann2004}.
The calculation of SM (flagged as \texttt{LGM\_TOT\_P50}) relies on theoretical models of stellar populations, following the methodology described by \cite{Kauffmann2003}. These computations assume the application of a Kroupa initial mass function (IMF; \citealt{Kroupa2001}).
Metallicity determination (flagged as \texttt{OH\_P50}) is based on the gas-phase oxygen abundance, expressed as $ 12 + \mathrm{log (O/H)} $, where the ratio $ \mathrm{O/H} $ signifies the abundance of oxygen relative to hydrogen.
Oxygen abundances are only computed for objects classified as "Star Forming" \citep{Tremonti2004, Brinchmann2004}. 

The Wide-field Infrared Survey Explorer (WISE; \citealt{Wright2010}) embarked on 
an all-sky survey project, focussing on the mid-infrared spectra with photometry in four filters at
$W1$ ($\lambda = 3.4$ $\mu$m), $W2$ ($\lambda = 4.6$ $\mu$m), $W3$ ($\lambda = 12$ $\mu$m),
$W4$ ($\lambda = 22$ $\mu$m). 
This endeavor led to the observation of an extensive array of celestial objects, generating over a million images.
AllWISE improves upon WISE by achieving deeper imaging, enhancing source detection, delivering superior photometric sensitivity and astrometric precision.
We adopt three magnitudes which are converted from Vega to AB magnitude as $ m_{\rm AB} = m_{\rm Vega} + \Delta m $
where $ \Delta m $ is 2.699, 3.339, and 5.174 in $ W1 $, $ W2 $, and $ W3 $ bands, respectively \citep{Schindler2017}. 
Since SDSS filters operate on the AB magnitude system, this conversion ensures that the magnitudes used in our analysis are on the same scale as the SDSS data.
In our analysis, we rely on observational errors obtained from SDSS and AllWISE. For SDSS, we utilize the \texttt{petroMagErr\_\textit{?}} values, where `?' corresponds to the five optical magnitudes: $u$, $g$, $r$, $i$, and $z$. For AllWISE, we employ the \texttt{e\_W\textit{?}mag} values, where `?' represents the WISE bands $W1$, $W2$, and $W3$. 

In order to obtain optical and infrared photometric data, 
we conduct a cross-match between the SDSS and AllWISE catalogues, 
using a match radius of 3 arcseconds \citep{Su2013}. This process 
results in a dataset comprising 526~409 galaxies. The machine learning algorithms 
adopt optical and IR features as inputs and are trained using output data 
from the SDSS MPA-JHU DR8 catalogue, specifically spectra-extracted 
SFR, SM, and metallicity. Our study aims to bridge gaps in understanding and accurately predict crucial galaxy properties from photometric information.

The SDSS MPA-JHU DR8 catalogue incorporates a classification system denoted by 
Baldwin--Phillips--Terlevich (BPT) classes (indicated by BPTCLASS; \citealt{Baldwin1981}). 
This diagram categorizes galaxies based on their emission line ratios, 
specifically $ \rm{[OIII] /H\beta} $ and $ \rm{[NII] /H\alpha} $. 
Within the classification system of the MPA-JHU catalogue, Class 1 corresponds 
to star-forming galaxies, Class 2 denotes low signal-to-noise star-forming, 
Class 3 represents composite galaxies, and Class 4 is attributed 
to AGNs\footnote{Active Galactic Nuclei} 
(excluding LINERS\footnote{low-ionization nuclear emission-line regions}). 
Additionally, Class 5 signifies low signal-to-noise LINERS. 
Moreover, the category Class -1 is designed for galaxies 
unclassified in the BPT diagram, particularly passive galaxies 
lacking emission lines \citep{Brinchmann2004}.

For our purposes, we select only data with high quality from the MPA-JHU DR8 catalogue, data that is considered to be reliable (\texttt{RELIABLE} $\neq$ 0) and without warning flags in redshift (\texttt{Z\_WARNING} $=$ 0). In addition, we keep data points without missing values for total SFR, total SM, and metallicity (\texttt{SFR\_TOT\_P50} $\neq$ -9999, \texttt{LGM\_TOT\_P50} $\neq$ -9999, \texttt{OH\_P50} $\neq$ -9999), and select only data with redshift \texttt{Z} $>$ 0.

The dataset sizes are as follows: 129~351 data points for metallicity, 509~711 data points for SFR, and 523~713 data points for SM. The distribution of the data is shown in Fig. \ref{fig:distribution} from left to right for SFR, SM, and metallicity, respectively.

To better understand the precision of our measurements, we computed the average widths of the 68\% confidence intervals for each target parameter across the training dataset. The average width of the 68\% confidence interval for the SFR parameter was found to be 0.551 dex. Similarly, for the SM and metallicity parameters, the average widths of the 68\% confidence intervals were 0.105 dex and 0.039 dex, respectively.

\begin{figure*}
\centering
        \begin{subfigure}[b]{0.33\textwidth}
                \centering
                \includegraphics[width=\linewidth]{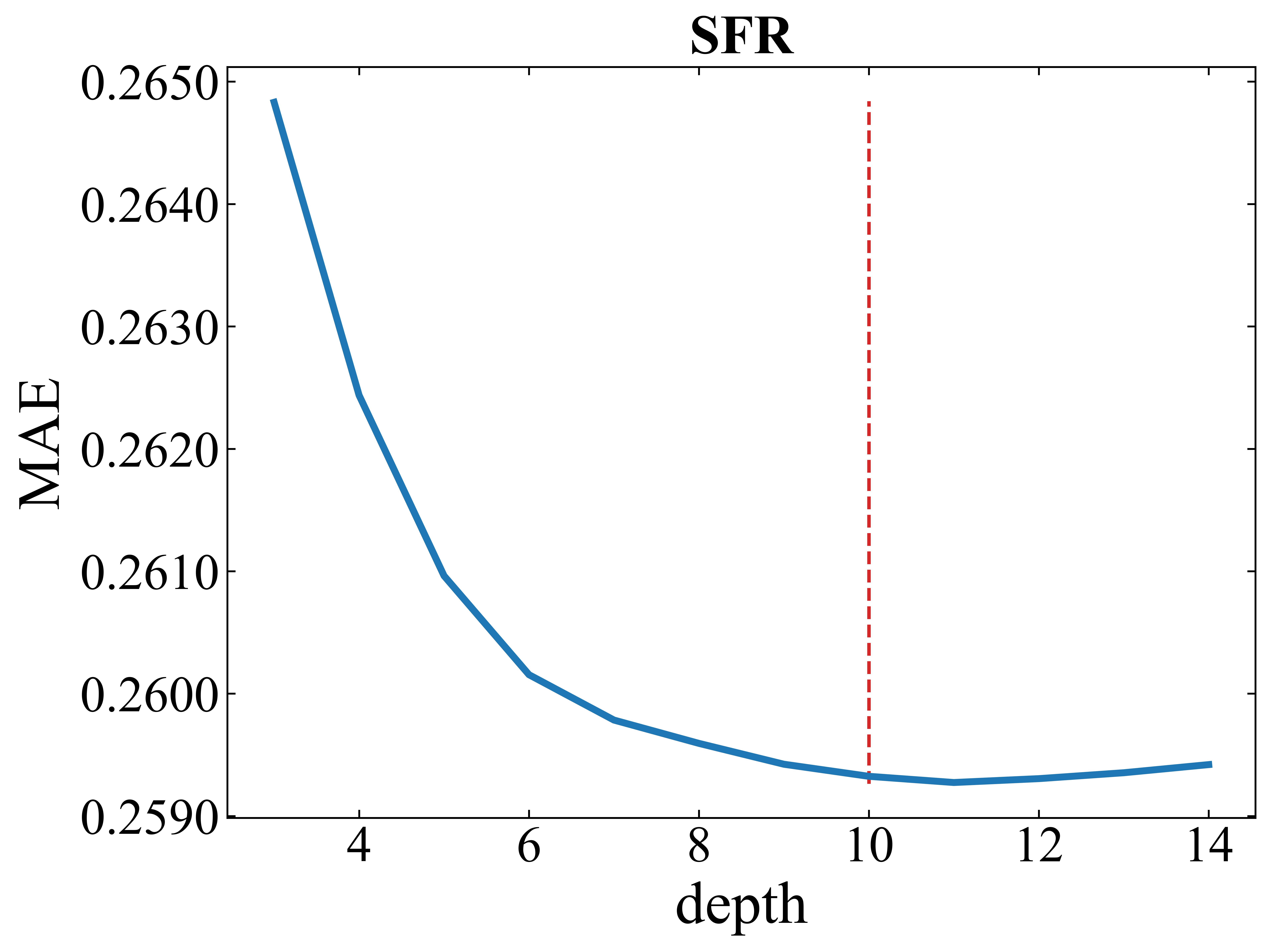}
                \label{fig:QSO}
        \end{subfigure}\hfill
        \begin{subfigure}[b]{0.33\textwidth}
                \centering
                \includegraphics[width=\linewidth]{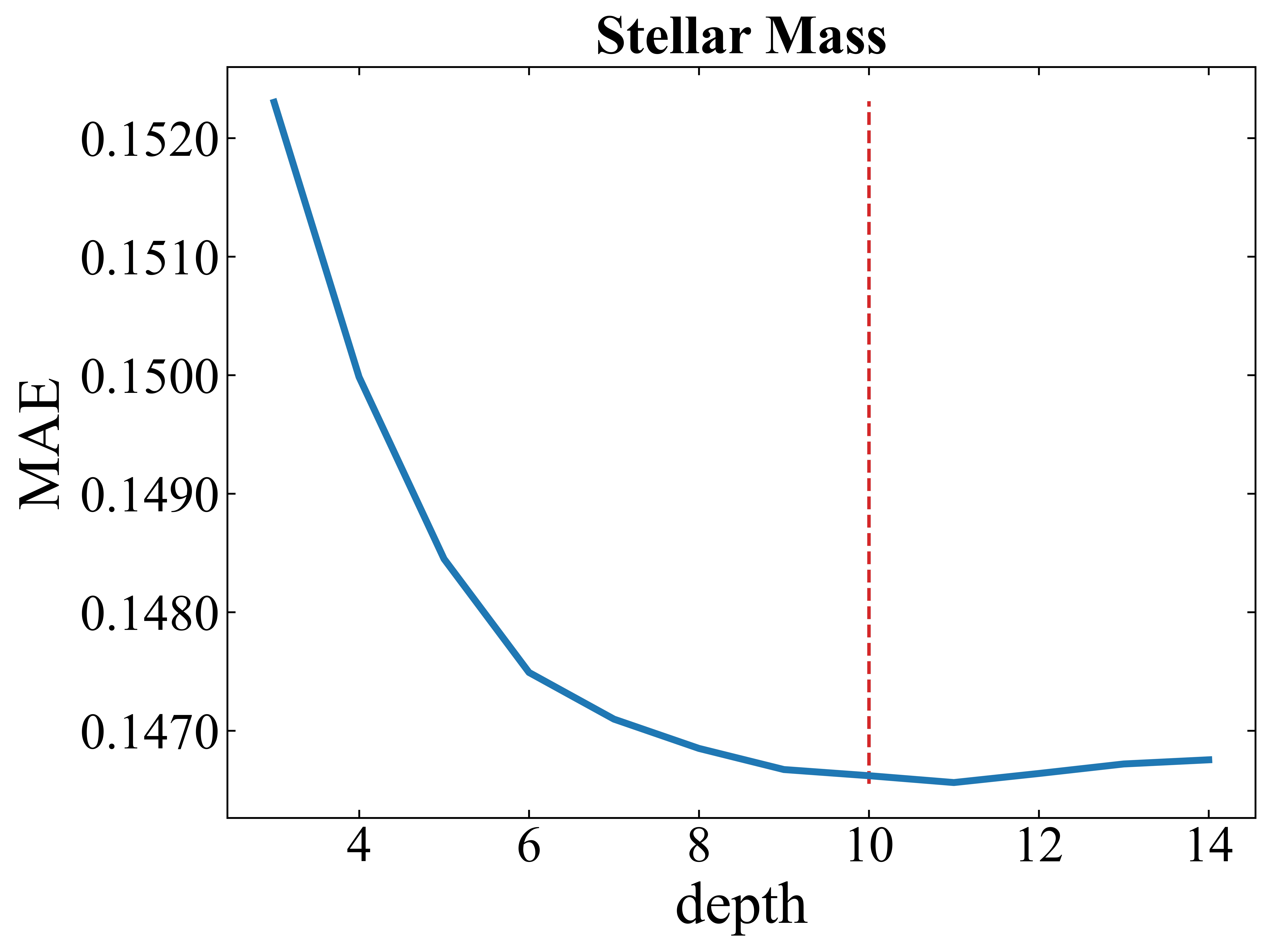}
                \label{fig:NG}
        \end{subfigure}\hfill
        \begin{subfigure}[b]{0.33\textwidth}
                \centering
                \includegraphics[width=\linewidth]{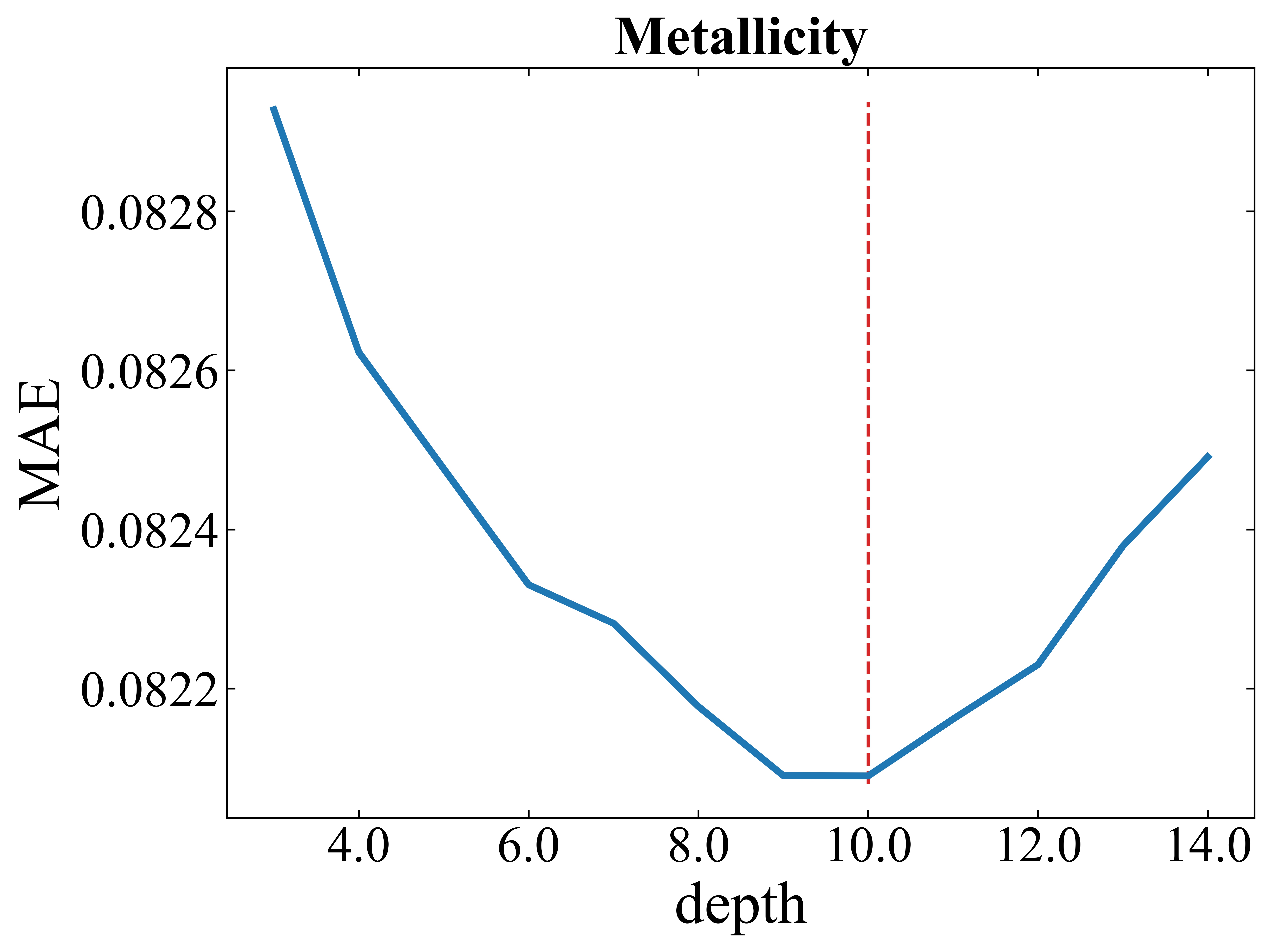}
                \label{fig:EG}
        \end{subfigure}\hfill
        \caption{Performance evaluation of CatBoost for different galaxy property predictions with varying `\textit{depth}' based on mean absolute error (MAE) for SFR (left panel), SM (middle panel), and metallicity (right panel). The results presented are the average performance across 10-fold cross-validation.}\label{fig:depth}
\end{figure*}
\section{Method}\label{method}
This paper delves into evaluating the performance of two supervised ML models, CatBoost and WDNN, focussing on quantifying a variety of physical parameters within galaxies. To evaluate the accuracy of estimating these parameters, we undertake a comparative analysis of the two methodologies.

To determine the properties of galaxies using ML based on optical and IR photometric data, we conducted tests involving several models, including random forest, XGBoost, CatBoost, and various DL structures. We explored different neural network structures and architectures, and the final presented architecture emerged from hyperparameter optimization to ensure a well-tailored and efficient model. After a comprehensive optimization process, we identified both the WDNN model and CatBoost as the top performers with our dataset. Consequently, our discussions will primarily centre around the outcomes derived from these high-performing models.

Given the significantly smaller dataset for metallicity compared to those for SFR and SM, we employed two distinct models—one for predicting SFR and SM and another for predicting metallicity. 
The dataset size for the ML and WDNN models targeting SFR and SM was 509~198, while the dataset for metallicity encompassed 129~351 instances. 
We note that the entire prediction process is conducted in logarithmic space for consistency.
Initially, to fine-tune the hyperparameters of CatBoost, we first split our dataset into dedicated training and test subsets. We randomly allocated 80\% for training and reserved 20\% for testing in both cases. Subsequently, we applied $k$-fold cross-validation \citep{Kohavi1995} with $k=10$ exclusively on the training subset for hyperparameter tuning. This method ensures that the test data remains untouched until the final performance evaluation, preserving its independence.

For the neural network, we employed a slightly different strategy. The dataset was initially partitioned into two sets: 20\% for testing and 80\% for training and hyperparameter tuning. Within the training set, 10\% was further allocated for validation, leaving 70\% for the actual model training. Here the validation set serves as an independent subset for hyperparameter fine-tuning.
Importantly, to ensure better comparability, all results presented in the tables are derived from the same test dataset for both CatBoost and the WDNN.

\subsection{CatBoost}
CatBoost, a member of the Gradient Boosted Decision Trees (GBDT) family introduced by \cite{Friedman2001} and developed by \cite{Dorogush2018}, is a high-performance open-source algorithm that utilizes boosting techniques to construct resilient predictive models. 
In gradient boosting, an ensemble of weak learners, typically decision trees, is built sequentially. Each tree corrects the errors of the preceding ones, gradually improving the overall predictive performance. Specifically, the algorithm focuses on minimizing the residuals of the previous trees, enhancing the ability of the model to capture complex relationships in the data.

Moreover, CatBoost incorporates regularization techniques to prevent overfitting and enhance model generalizability. The algorithm also provides built-in functionalities for cross-validation, streamlining hyperparameter tuning and facilitating model performance assessment. 
The central hyperparameter, `\textit{depth}', plays a pivotal role in controlling the complexity of the decision trees. It determines the maximum depth of each tree in the ensemble. A higher depth allows the model to capture intricate patterns in the training data, but it also increases the risk of overfitting. On the other hand, a lower depth may lead to underfitting, where the model fails to capture essential details. Striking the right balance for the `\textit{depth}' parameter is crucial for achieving optimal predictive performance.
Notably, the default settings of CatBoost often yield strong performance, with optimization efforts typically focussed on adjusting the `\textit{depth}' parameter while retaining other parameters at their default values.

In our study, 10-fold cross-validation was conducted, and mean absolute error (MAE) values with varying `\textit{depth}' are illustrated in Fig. \ref{fig:depth}. Optimal performance was achieved with a `\textit{depth}' parameter set to 10, resulting in MAE values of about 0.259~dex, 0.147~dex, and 0.082~dex for SFR, SM, and metallicity, respectively.

\begin{figure*}
  \centering
  \subfloat{\includegraphics[width=0.5\linewidth, trim=0.9in 0.1in 0.9in 0.0in, clip]{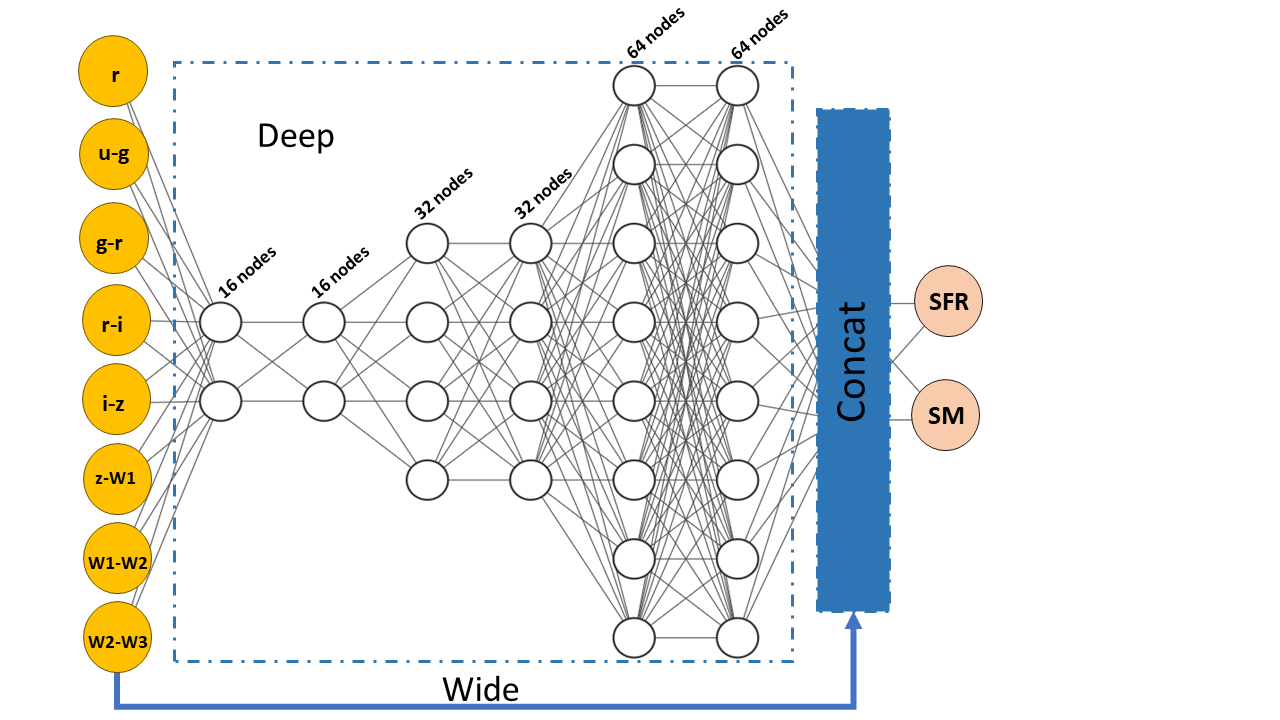}}\hfill 
  \subfloat{\includegraphics[width=0.5\linewidth, trim=0.9in 0.1in 0.9in 0.0in, clip]{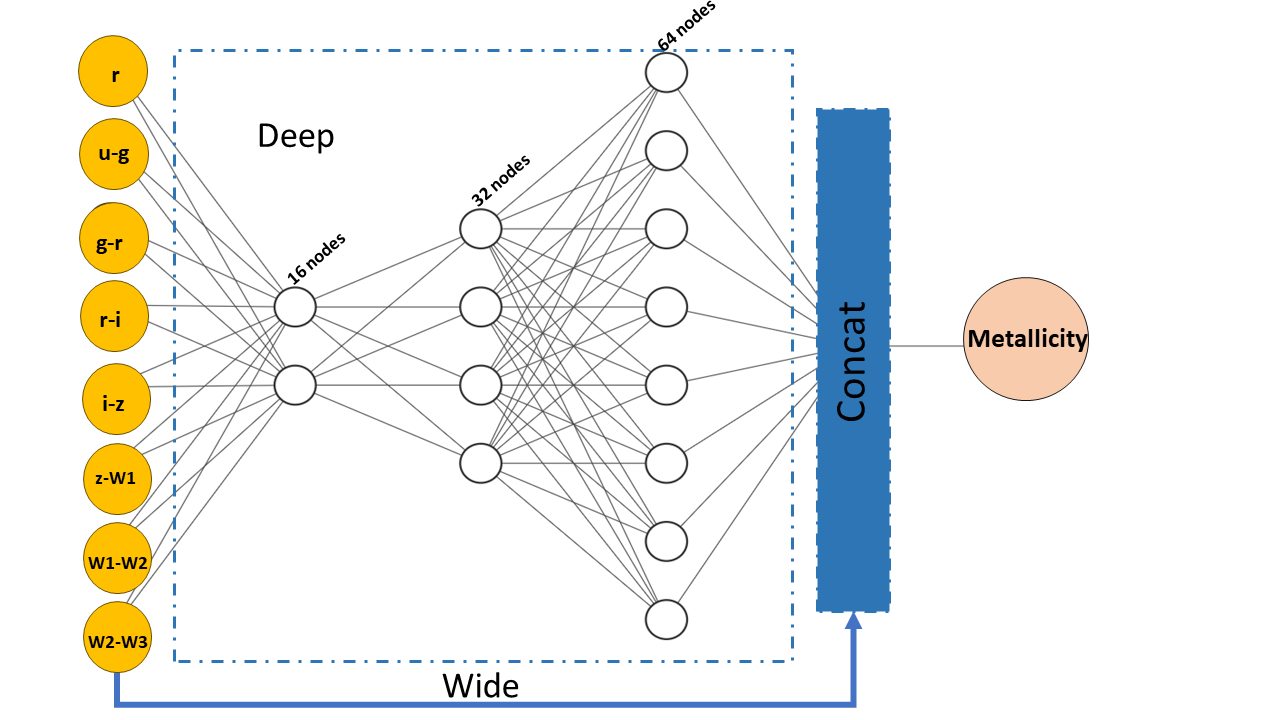}}\hfill
  \caption{The schematic structure of the WDNN models utilized in this study. The input features consist of a combination of 8 optical and IR magnitudes and colours. The left panel outputs SFR and SM, while the right panel focuses on predicting metallicity.}
  \label{fig:DNN}
\end{figure*}

\subsection{Deep learning}
In our study, we employed the `Wide and Deep' model \citep{Cheng2016} for deep learning using Keras \citep{Chollet2015}. Unlike the conventional neural networks that require data to pass through each layer, potentially obscuring simpler patterns, the Wide and Deep Neural Network (WDNN) allows for non-sequential processing. This capability enables the network to simultaneously comprehend complex patterns (via the deep path) and simple rules (via the short path). This unique design fosters a more comprehensive understanding of the complexities of the data compared to the sequential processing of traditional neural networks.

The architecture of the two WDNNs used in our study is shown in Fig. \ref{fig:DNN}. The model displayed in the left panel aims to predict SFR and SM, while the model in the right panel focuses solely on predicting metallicity. The left model consists of six hidden layers, each structured with neurons arranged as [16,16,32,32,64,64] and utilizing Rectified Linear Unit (ReLU) activation functions. A Concatenate layer combines the input and output of the sixth hidden layer, resulting in the output layer comprising two neurons (SFR and SM) with a linear activation function.

The right model in Fig. \ref{fig:DNN}, tailored towards the prediction of metallicity, entails
three hidden layers with [16,32,64] neurons using the Exponential Linear Unit (ELU) activation function. The third hidden layer of this model is then concatenated with the input and output of the Concatenate layer. The output layer here consists of a single neuron (metallicity) with a linear activation function.

Different from the CatBoost model, WDNN entails refining various hyperparameters such as the number of neurons, layers, activation functions, and kernel initializers. 
These hyperparameters undergo careful adjustments during training, spanning up to 1000 epochs with the Adam optimizer, and the training process employs the mean square error (MSE) as the loss function.
Training continues until the difference between loss values of two consecutive epochs is below the machine precision level.
 
Furthermore, we integrate the ReduceLROnPlateau technique, which dynamically modifies the learning rate based on the performance of the model on a dedicated validation set.
Specifically, if the performance metric remains constant for several consecutive epochs during training, we systematically decrease the learning rate by a specified amount. This systematic adjustment not only improves model performance but also enhances its adaptability to the complexities of the data. Lowering the learning rate helps the model navigate the optimization landscape more effectively, potentially aiding in overcoming local minima and expediting convergence or fine-tuning. 

\subsection{Evaluation metrics}
We evaluate the accuracy and efficacy of our regression models using metrics such as the residual between the true values (targets) ($X_{\rm t}$) of the galaxy properties obtained by SED fitting through spectra and the predicted values ($X_{\rm p}$) by our regression models for each property in our dataset as
\begin{equation}
\Delta X = X_{\rm p} - X_{\rm t}
\end{equation}
where $X$ represents any of the three galaxy properties: SFR, SM, and metallicity.
These differences serve as quantitative indicators of how well our models approximate the true galaxy properties.

The mean squared error (MSE)  measures the average squared difference between the predicted values and the true values as
\begin{equation}
\mathrm{MSE} = \frac{1}{n} \sum_{i=0}^{n-1} (X_{{\rm p},i} - X_{{\rm t},i})^{2}
\end{equation}
where ${X_{{\rm t},i}}$ is the true value for the $i$th galaxy, ${X_{{\rm p},i}}$ is the predicted value for the $i$th galaxy, and $n$ is the sample size.
The root mean squared error (RMSE) is the square root of MSE and provides an interpretable scale similar to the target variable as
\begin{equation}
\mathrm{RMSE = \sqrt{\mathrm{MSE}}}
\end{equation}

The normalised root mean squared error (NRMSE) is a normalised version of the RMSE and calculated by dividing the RMSE by the range (the difference between the maximum and minimum values) of the true target values.

\begin{equation}
\text{NRMSE} = \frac{\text{RMSE}}{\text{Range of true values}}
\end{equation}

The mean absolute error (MAE) measures the average absolute difference between the predicted values and the true values. It is less sensitive to outliers than MSE. 
\begin{equation}
\mathrm{MAE} = \frac{1}{n} \sum_{i=0}^{n-1} \mid X_{{\rm p},i} - X_{{\rm t},i} \mid
\end{equation}

The standard deviation of the residuals is calculated as
\begin{equation}\label{eq:sigma}
\sigma = \sqrt{\frac{1}{n} \sum_{i=0}^{n-1} 
(\Delta X_{i} - \overline{\Delta X_{i}})^2}
\end{equation}

Bias is defined as the average separation between prediction and true values as
\begin{equation}
\mathrm{bias} = \langle X_{\rm p} - X_{\rm t} \rangle
\end{equation}

In our study on estimating galaxy properties, $\eta$ represents 
the percentage of catastrophic outliers. We apply a criterion of 
$\mid\Delta X\mid > 3\sigma $, where data points with residuals 
exceeding three times the standard deviation are considered potential catastrophic outliers.
\begin{figure*}
\centering
        \begin{subfigure}[b]{0.5\textwidth}
                \centering
                \includegraphics[width=\linewidth]{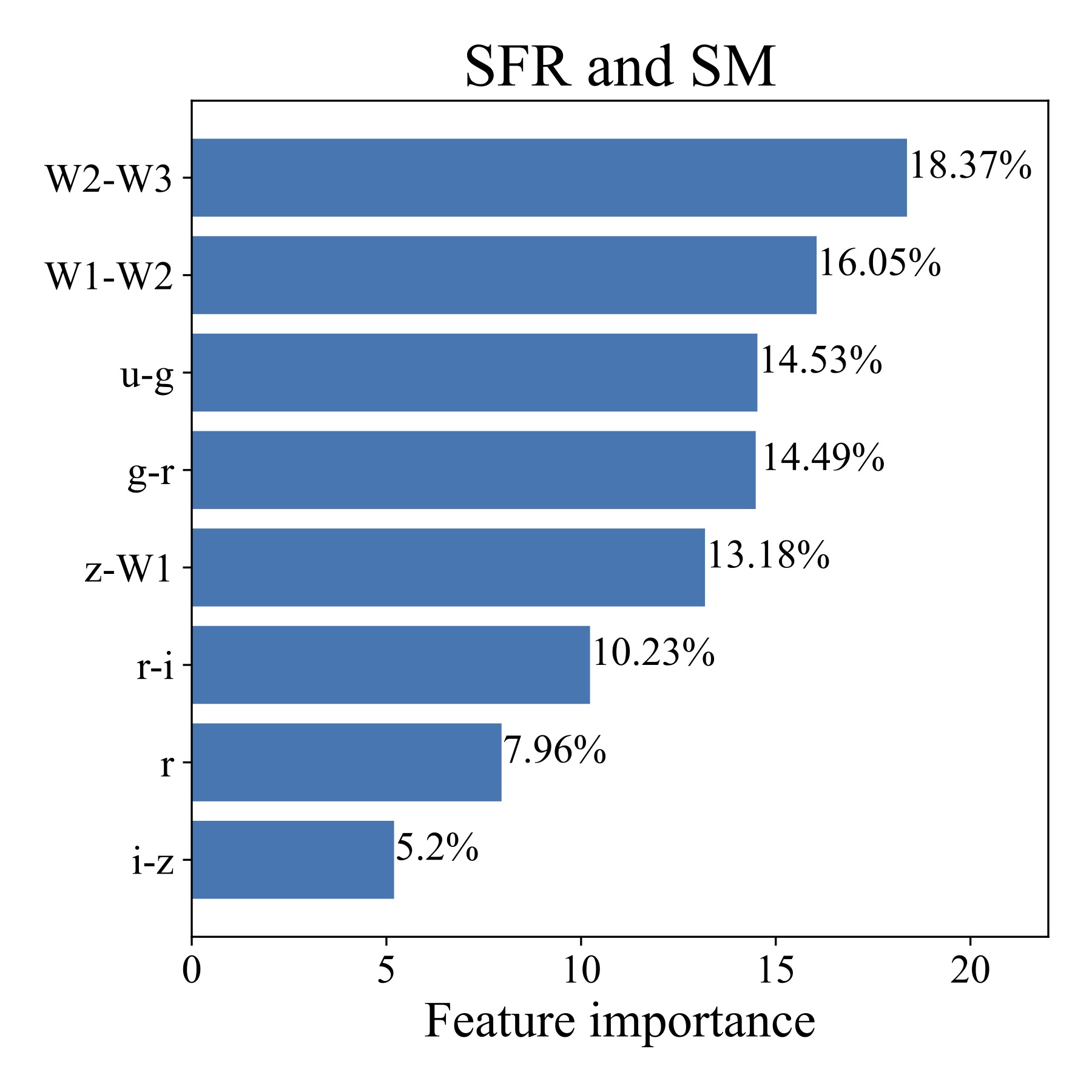}
                \label{fig:feature_SFR}
        \end{subfigure}\hfill
        \begin{subfigure}[b]{0.5\textwidth}
                \centering
                \includegraphics[width=\linewidth]{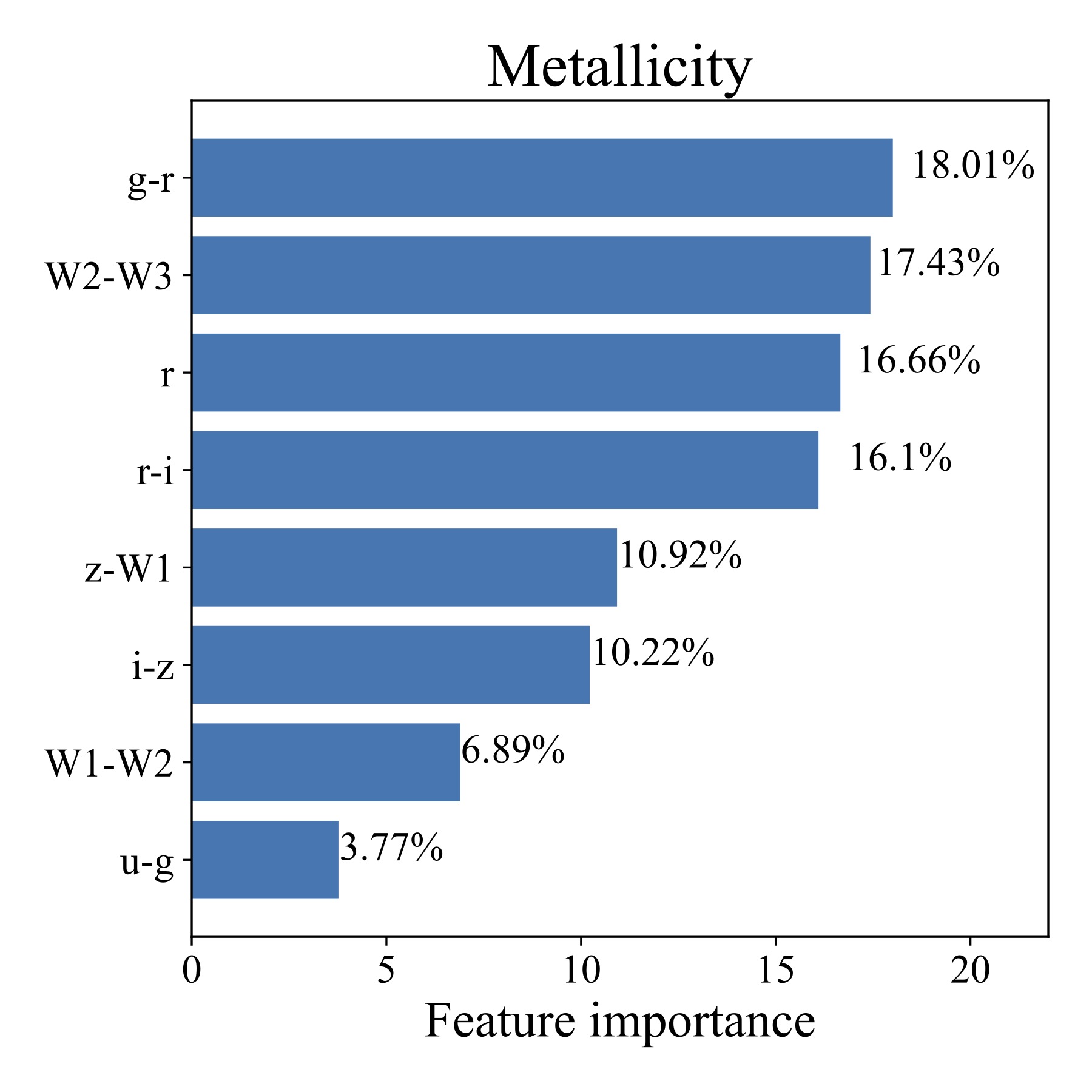}
                \label{fig:feature_M}
        \end{subfigure}\hfill
        \caption{Feature importance analysis by CatBoost: the left panel demonstrates the significance of features for predicting SFR and SM, while the right panel provides insights into feature importance for predicting metallicity.}\label{fig:feature_importance}
\end{figure*}
\section{Machine learning model}\label{model}
\subsection{Feature selection}
The careful choice of input features plays an important role in achieving the optimal performance of an ML model.
In our experiments, a set of 8 optical and IR magnitudes and colours, including $r, u-g, g-r, r-i, i-z, z-W1, W1-W2$, and $W2-W3$, from previous studies \citep{Bonjean2019, Li2023}, was employed as input features for training the models. 

The importance of a feature in addressing a learning problem is evaluated based on its contribution. 
By employing CatBoost, we assessed the importance scores for each feature and organised them in descending order based on their significance, as depicted in Fig. \ref{fig:feature_importance}. The CatBoost importance score, derived from this evaluation, quantifies the contribution of each feature to the predictive performance of the model. A higher CatBoost importance score signifies a more significant role of a feature in predicting the target variable within the model. Importantly, these feature importance scores provided by CatBoost consider not only linear relationships but also nonlinear contributions of features to the predictive capabilities of the model. Evidently from Fig. \ref{fig:feature_importance}, the most influential features for predicting SFR and SM are IR colours $W2-W3$ and $W1-W2$, with scores of 18.37\% and 16.05\%, respectively. In contrast, for predicting metallicity, the key features are both optical and IR colours $g-r$ and $W2-W3$, with scores of 18.01\% and 17.43\%, respectively.


\begin{table*}
\small
\centering
 \begin{threeparttable}
\caption{Comparative evaluation metrics for ML models (CatBoost and WDNN) on the physical parameters SFR, SM, and metallicity.}
\label{tab:performance}
\begin{tabular}{lcccccccccc}
\toprule
\multirow{2}{*}{Parameter} & \multicolumn{5}{c}{\textbf{CatBoost}} & \multicolumn{5}{c}{\textbf{WDNN}\hspace{1em}} \\
\cmidrule(lr){2-6} \cmidrule(lr){7-11}
& \textbf{RMSE} & \textbf{NRMSE} & \textbf{MAE} & \textbf{Bias} & $\boldsymbol{\eta}$ & \textbf{RMSE} & \textbf{NRMSE} & \textbf{MAE} & \textbf{Bias} & $\boldsymbol{\eta}$ \\
\midrule
SFR & 0.349 & 0.040 & 0.259  & 0.002 & 0.011 & 0.351 & 0.040 & 0.260 & 0.001 & 0.011 \\
SM & 0.209 & 0.034 & 0.146 & 0.001 & 0.013 &  0.210  & 0.034 & 0.147 & 0.001 & 0.013\\
Metallicity & 0.109 & 0.064 & 0.082 & 5.354e-05 & 0.011 & 0.110 & 0.065 & 0.083 & 0.002 & 0.010 \\
\bottomrule
\end{tabular}
 \begin{tablenotes}
    \item \scriptsize \textbf{Note:} The SFR is measured in units of $ \rm{log(SFR[M_{\odot}yr^{-1}])} $, SM is also measured in units of $ \rm{log(M^{\star}[M_{\odot}])} $, and metallicity is represented as $\rm{12+log(O/H)}$. All numerical values presented in the tables have been rounded to the third decimal place for clarity.
    \end{tablenotes}
    \end{threeparttable}
\end{table*}

\begin{figure*}
  \centering
  \subfloat{\includegraphics[width=0.5\linewidth]{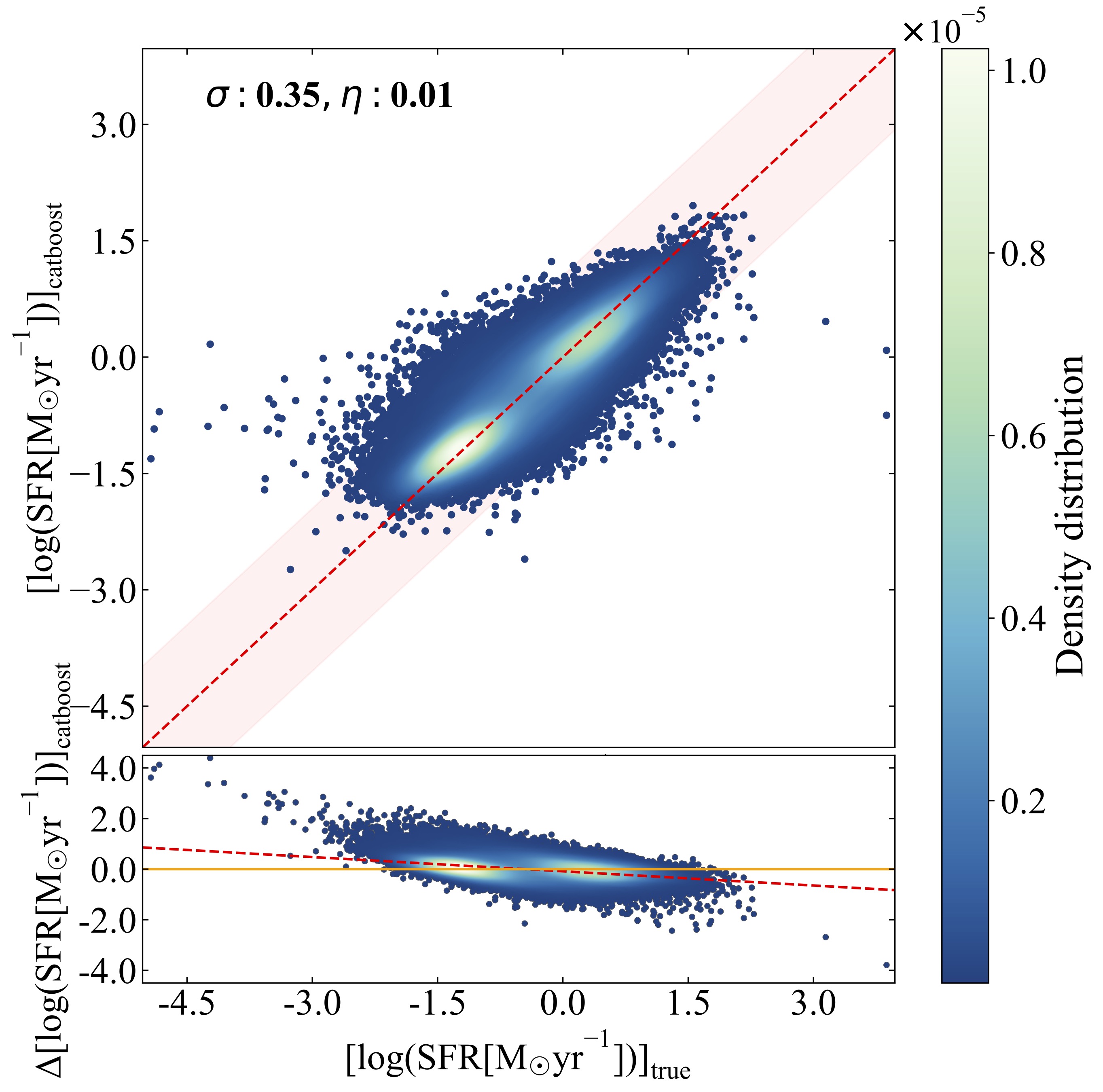}}\hfill
  \subfloat{\includegraphics[width=0.5\linewidth]{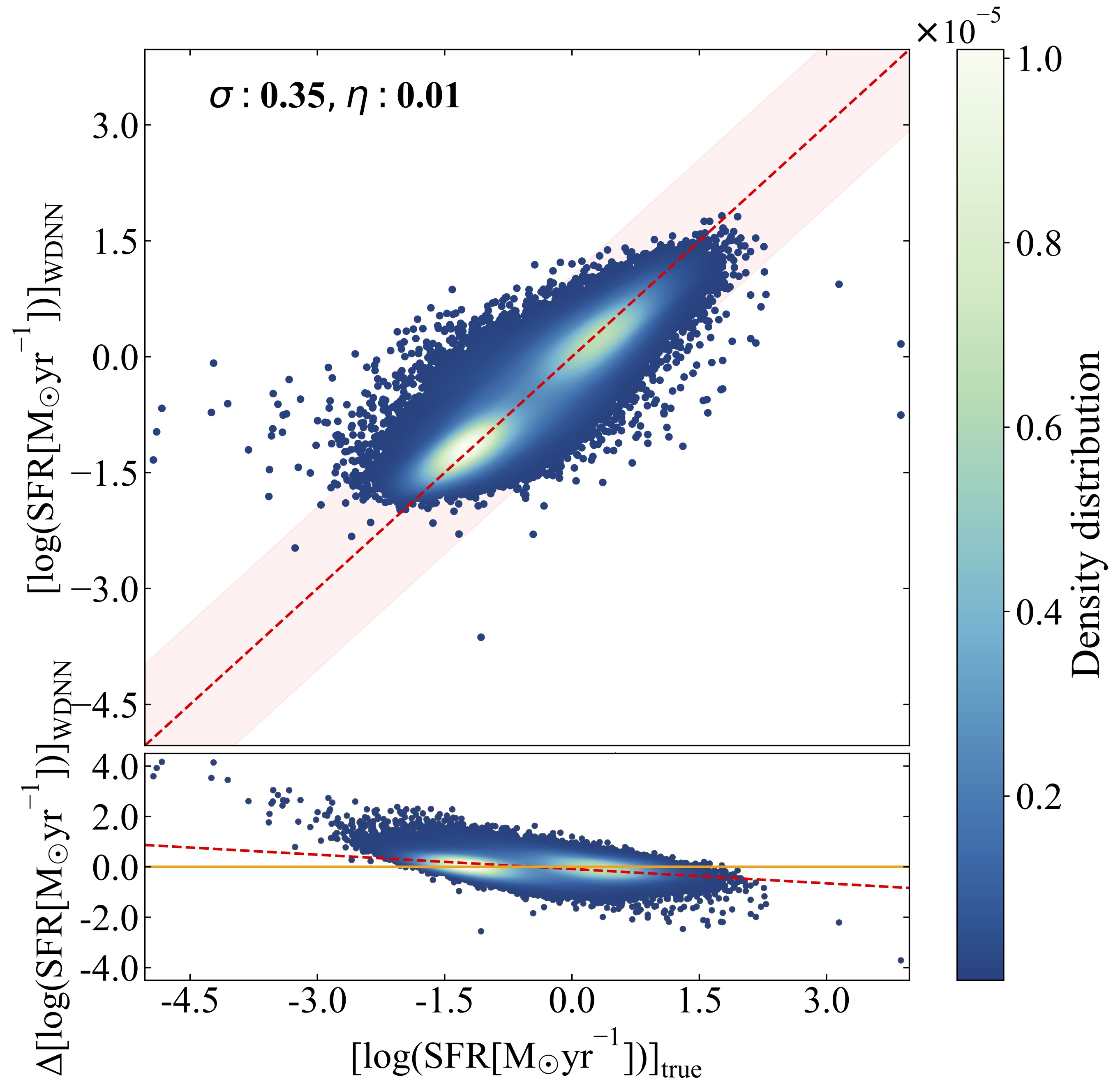}}\hfill
  \subfloat{\includegraphics[width=0.5\linewidth]{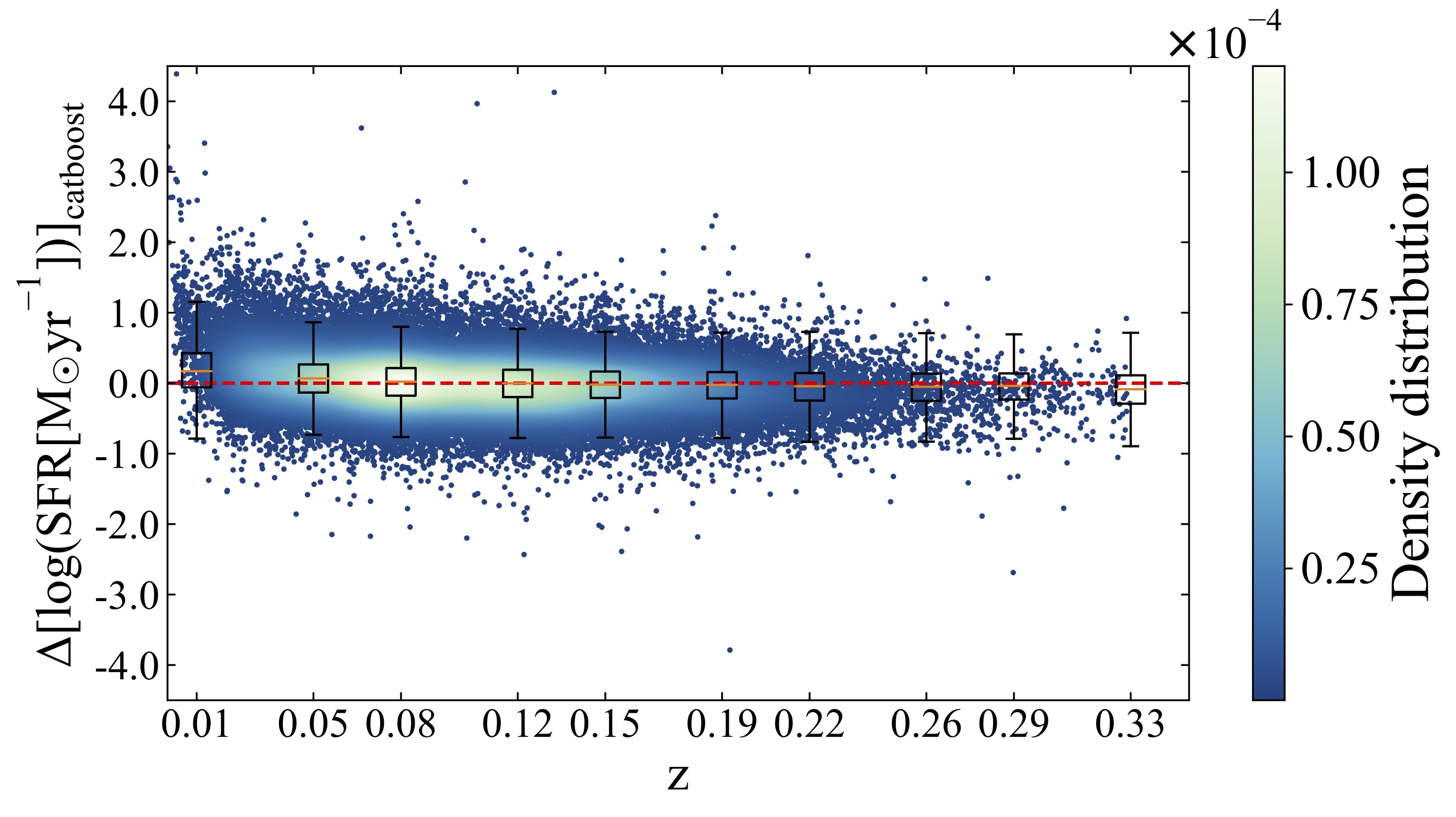}}\hfill
  \subfloat{\includegraphics[width=0.5\linewidth]{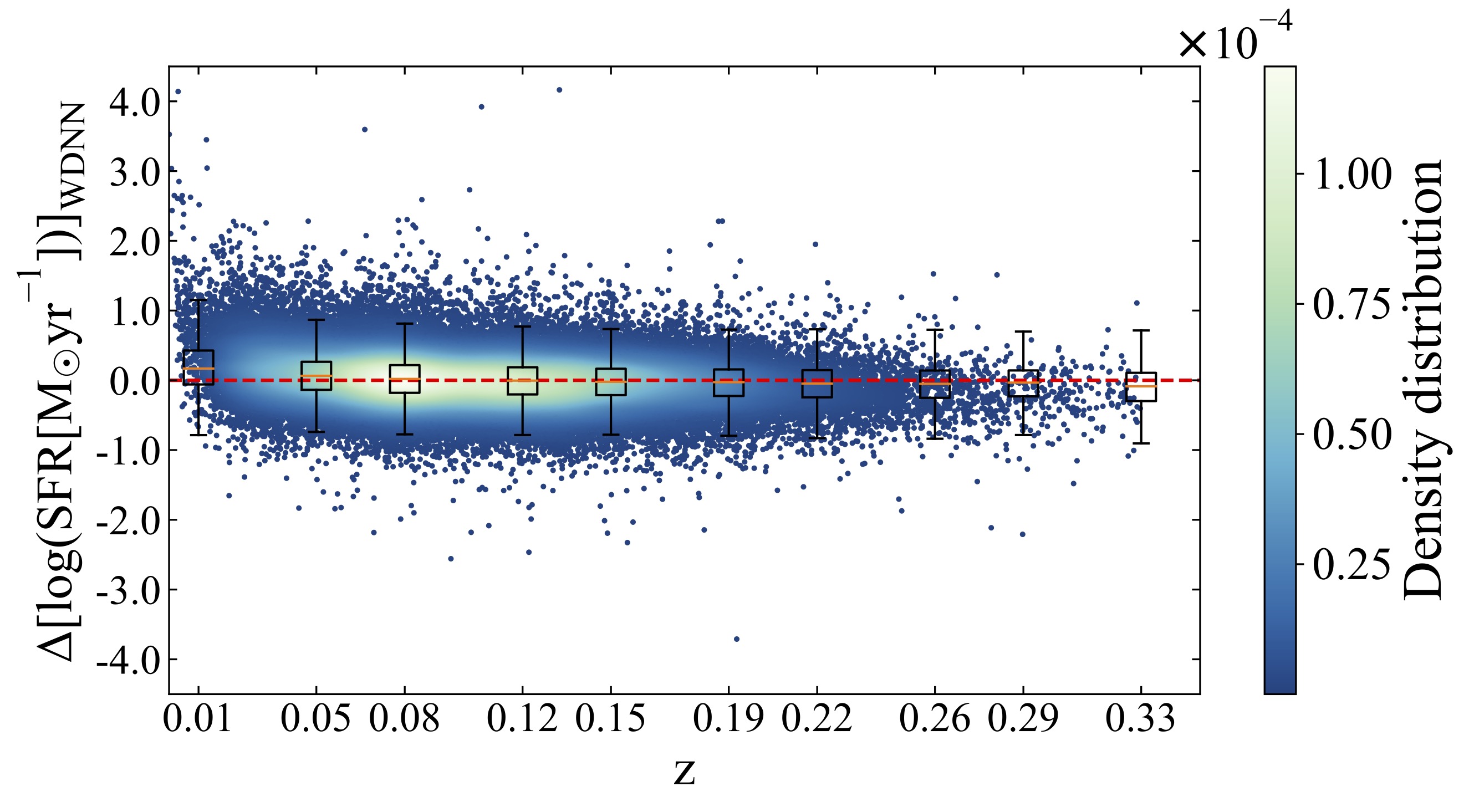}}\hfill
  \caption{Results of CatBoost (left panels) and WDNN (right panels) on the test sample (20\% of the entire dataset):
The upper row subplots present a comparison of SFR estimates from the ML model with SFR values obtained from the SDSS MPA-JHU DR8 catalogue, displaying errors in the results obtained by ML models about the true values at the bottom. The pink-shaded region indicates a $3\sigma $ scatter of the SFR errors, with any data point falling outside these limits considered an outlier. 
In the bottom row, box plots illustrate errors across redshift values. 
In the residual plots, the dashed lines illustrate the best fit through the scatter plots.
}
  \label{fig:SFR}
\end{figure*}
\begin{figure*}
  \centering
  \subfloat{\includegraphics[width=0.5\linewidth]{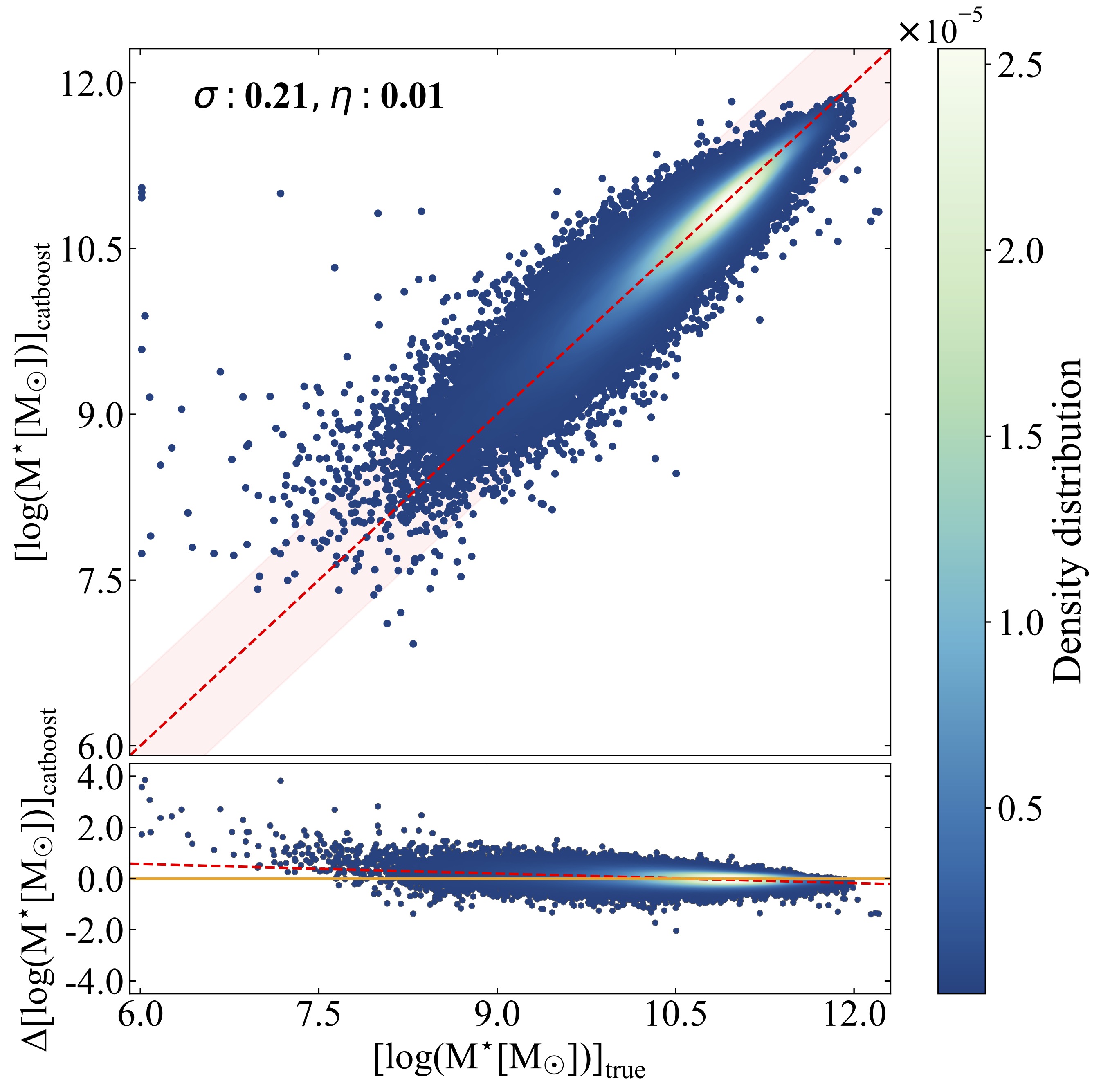}}\hfill
  \subfloat{\includegraphics[width=0.5\linewidth]{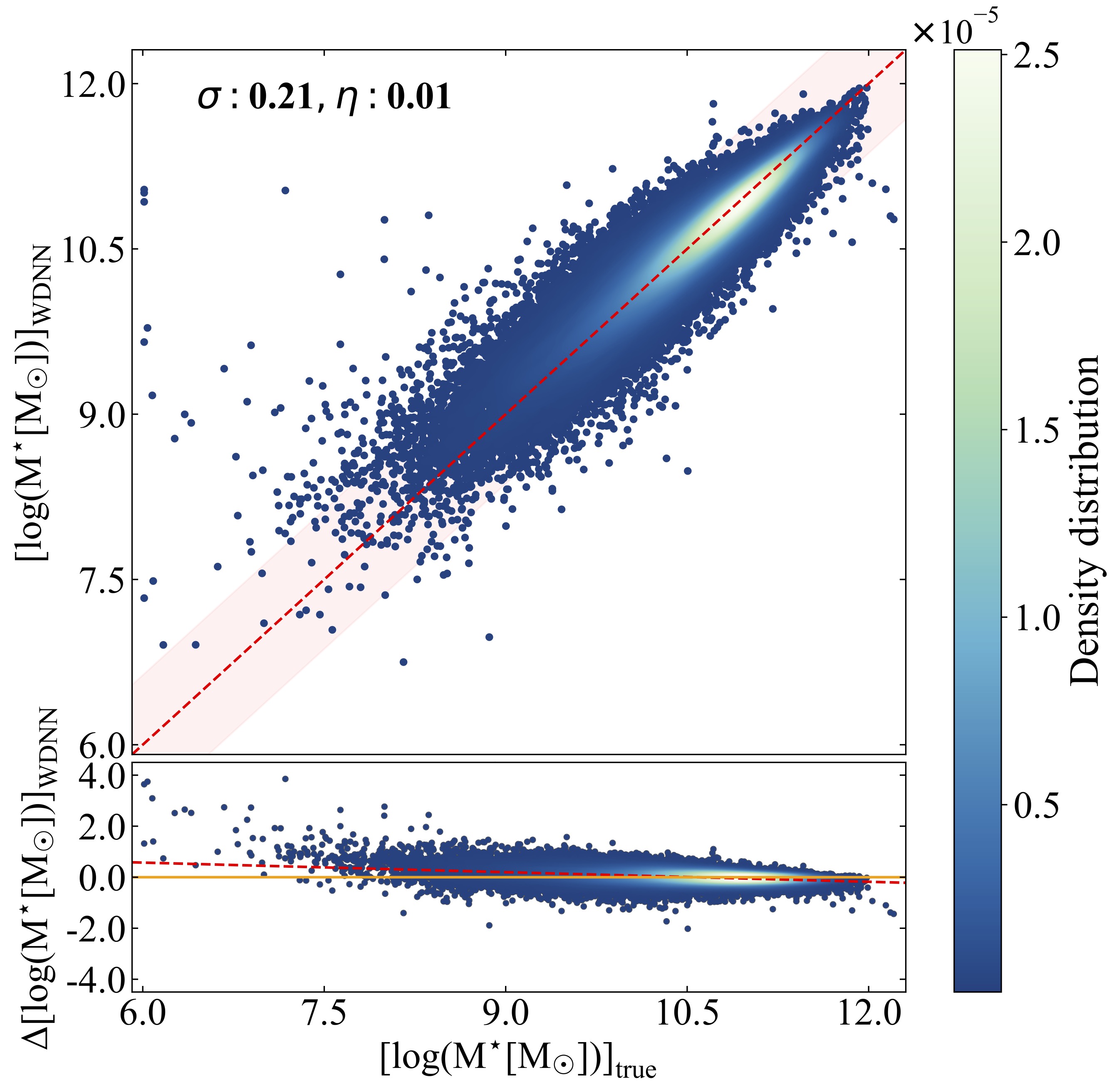}}\hfill
  \subfloat{\includegraphics[width=0.5\linewidth]{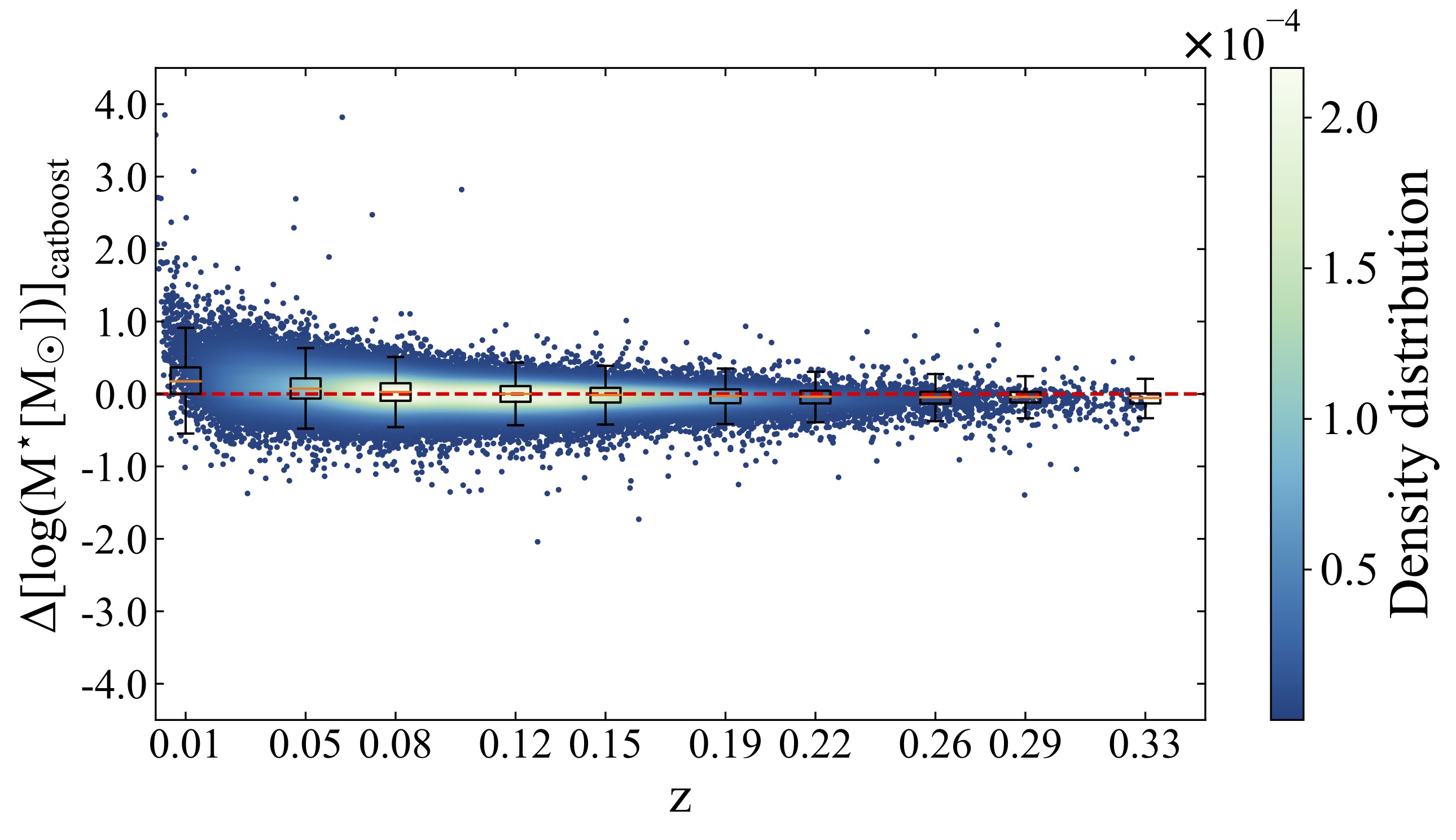}}\hfill
  \subfloat{\includegraphics[width=0.5\linewidth]{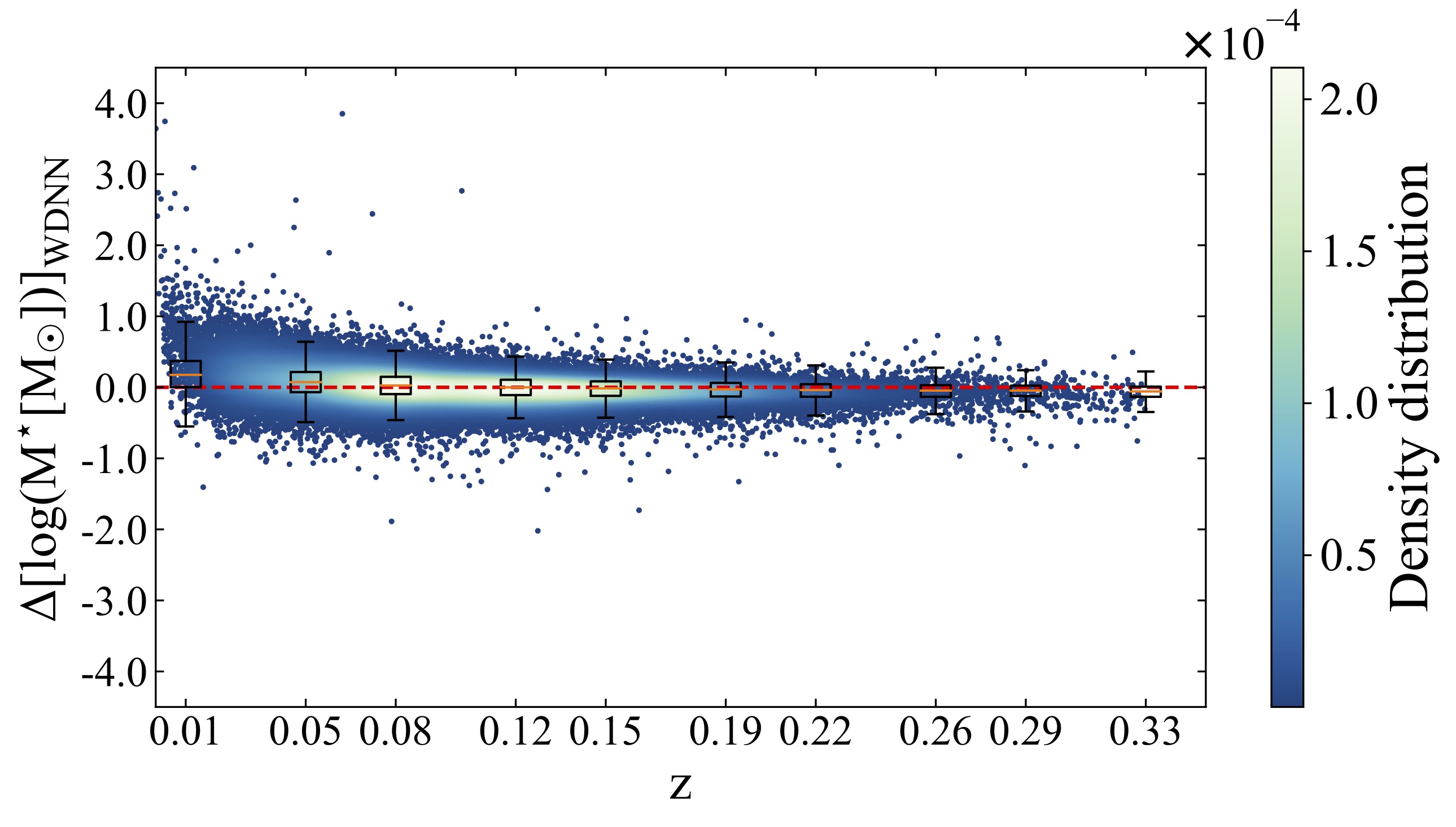}}\hfill
  \caption{Results of CatBoost (left panels) and WDNN (right panels) on the test sample (20\% of the entire dataset):
The upper row subplots present a comparison of SM estimates from the ML model with SM values obtained from the SDSS MPA-JHU DR8 catalogue, displaying errors in the results obtained by ML models about the true values at the bottom. The pink-shaded region indicates a $ 3\sigma $ scatter of the SM errors, with any data point falling outside these limits considered an outlier. In the bottom row, box plots illustrate errors across redshift values. 
In the residual plots, the dashed lines illustrate the best fit through the scatter plots.}
  \label{fig:SM}
\end{figure*}
\begin{figure*}
  \centering
  \subfloat{\includegraphics[width=0.5\linewidth]{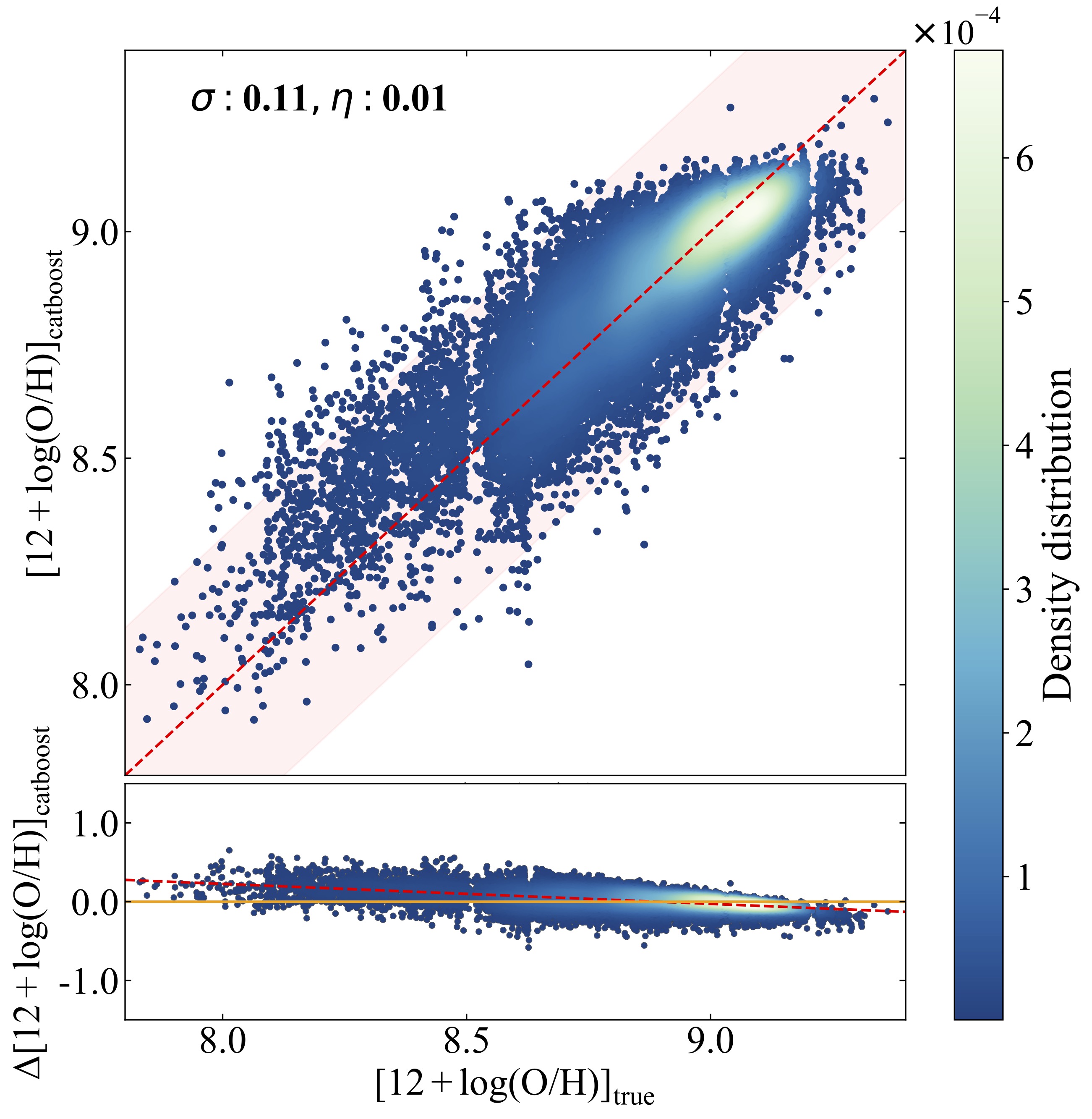}}\hfill
  \subfloat{\includegraphics[width=0.5\linewidth]{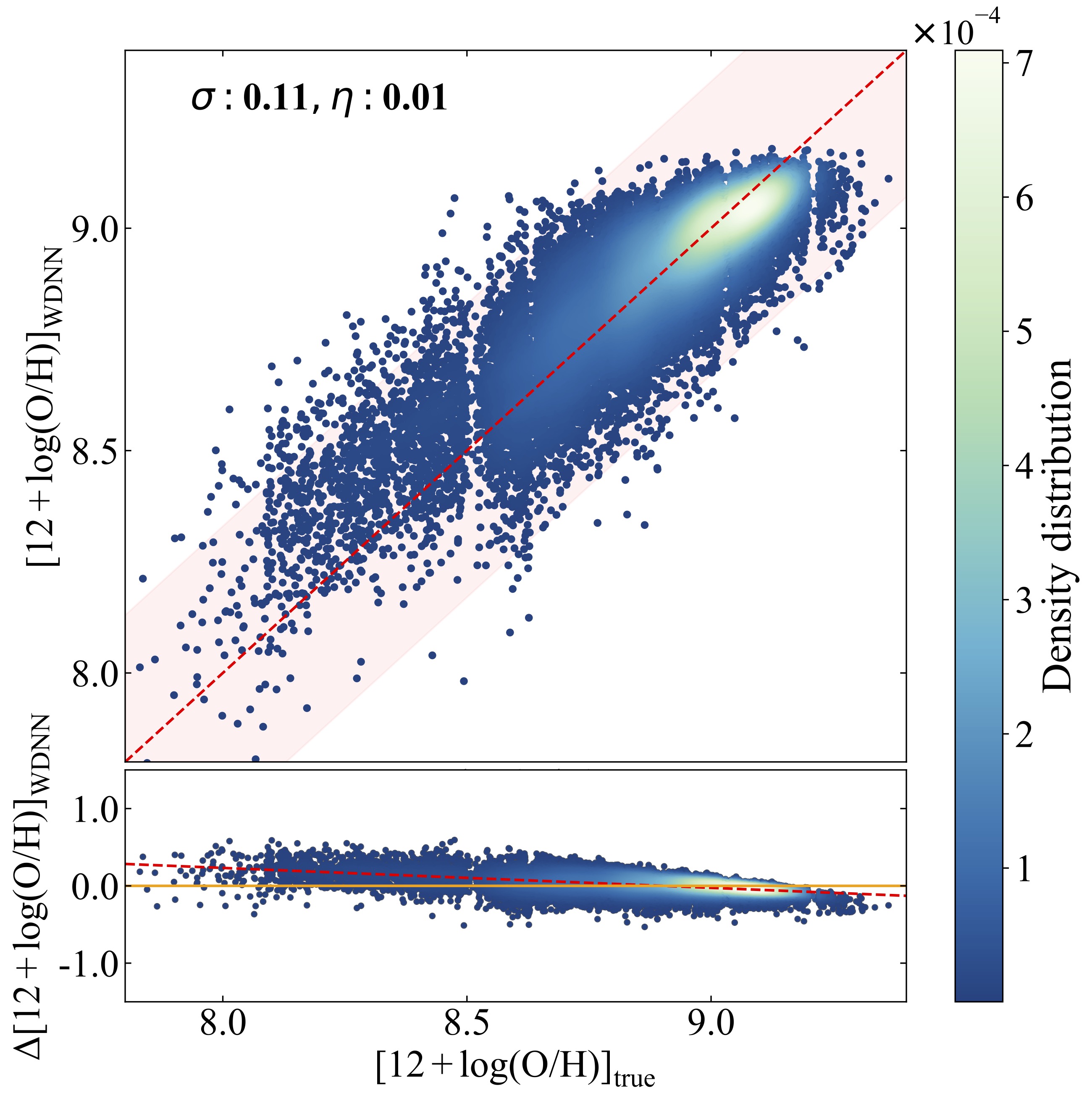}}\hfill
  \subfloat{\includegraphics[width=0.5\linewidth]{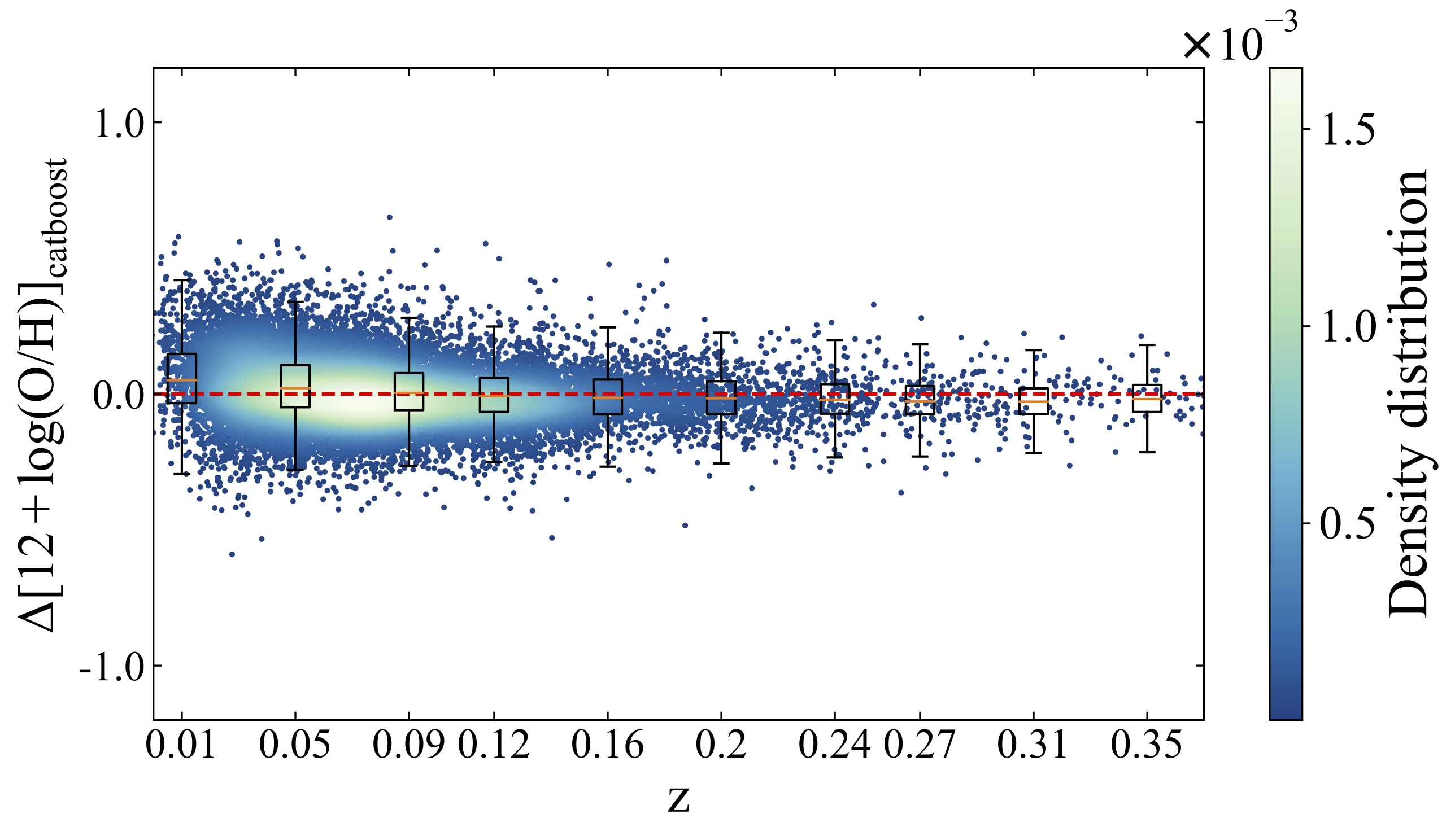}}\hfill
  \subfloat{\includegraphics[width=0.5\linewidth]{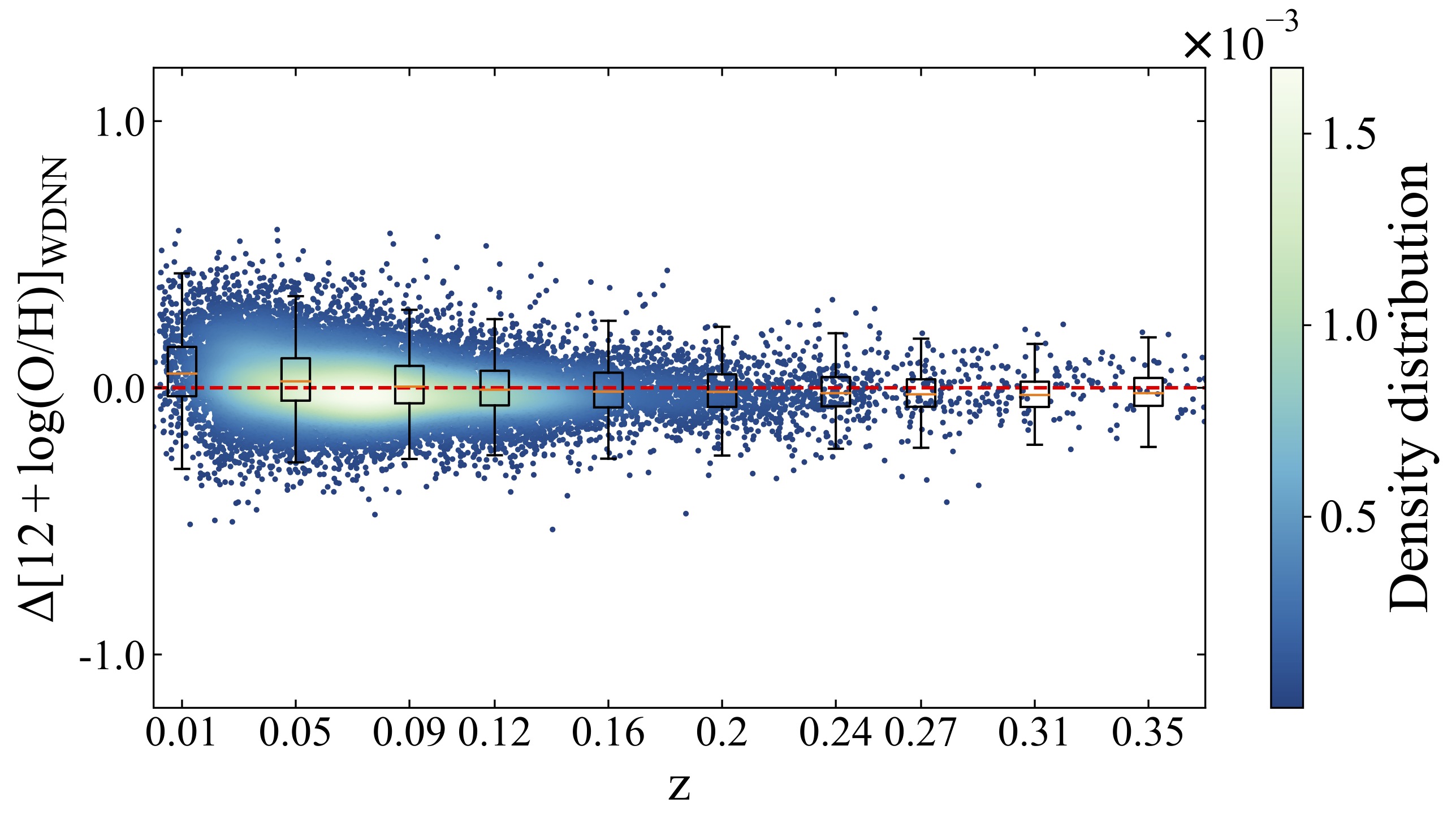}}\hfill
  \caption{Results of CatBoost (left panels) and WDNN (right panels) on the test sample (20\% of the entire dataset):
The upper row subplots present a comparison of metallicity estimates from the ML model with metallicity values obtained from the SDSS MPA-JHU DR8 catalogue, displaying errors in the results obtained by ML models about the true values at the bottom. The pink-shaded region indicates a $ 3\sigma $ scatter of the metallicity errors, with any data point falling outside these limits considered an outlier. In the bottom row, box plots illustrate errors across redshift values. In the residual plots, the dashed lines illustrate the best fit through the scatter plots.}
  \label{fig:M}
\end{figure*}
\subsection{Results}
\subsubsection{Performance of CatBoost and WDNN}\label{sec:galaxies}
We trained CatBoost and WDNN models on
the training set to estimate SFR, SM,
and metallicity on the test set. 
The assessment of the models was performed by comparing the results with the true values 
from the SDSS MPA-JHU DR8 catalogue. For a comprehensive evaluation, we employed 
a range of key metrics outlined in Table \ref{tab:performance}, 
including RMSE, NRMSE, MAE, bias, and the proportion of catastrophic outliers $ \eta $.

Our methodology involves a dual-model approach to evaluate CatBoost and WDNN. Each model is examined in two separate configurations, each focussing on different output targets. The first model predicts SFR and SM, while the second model predicts metallicity.
The upper row subplots in Figs. \ref{fig:SFR}, \ref{fig:SM}, and \ref{fig:M} represent the disparities between the ML estimated values and the true data. Below these subplots, the errors are plotted about the true values, providing a visual representation of how the residuals fluctuate concerning the actual data points. 
The consistent performance of both ML models is clearly demonstrated by the strong agreement between the true values and predictions by both ML algorithms. 
Comparing the performance of CatBoost (left panels) and WDNN (right panels), we find a comparable result.

The results presented in Table \ref{tab:performance} further reinforce that the performances of CatBoost and WDNN are comparable across all metrics.
Minimal differences are observed, with CatBoost generally demonstrating slightly lower RMSE values than the WDNN model, indicating marginal advantages, typically around 0.001.

The NRMSEs indicate that the models exhibit high accuracy in predicting galaxy parameters, as evidenced by their small values. Specifically, the NRMSE for SFR is 0.040, for SM it is 0.034, and for metallicity, it is 0.064. Analyzing these values, we observe that the highest NRMSE is associated with metallicity. Therefore, it can be inferred that metallicity poses a greater challenge for prediction compared to the other two parameters. This could imply that the features or factors influencing metallicity in galaxies are more complex or less straightforward for the model to capture accurately. It may also indicate that there is greater variability or uncertainty in the data related to metallicity.
Additionally, the comparatively worse performance of the metallicity prediction could be related to the fact that the training data for these models is about a factor 5 smaller than that for SFR and SM.

The bottom row subplots of Figs. \ref{fig:SFR}, \ref{fig:SM}, and \ref{fig:M} illustrate the comparison of model accuracy across different redshifts. These figures present both residual scatter plots and box plots plotted alongside redshift. Upon examination, 
we observe symmetrical box plots, indicating that the variability or spread in the residuals is consistent across the range of redshifts being considered.
However, we also noted that at lower redshifts, the whisker lengths in the box plots were significantly longer compared to higher redshifts. This observation hints at a greater variability or spread in the residuals at lower redshifts, potentially posing challenges for accurate prediction of galaxy properties within this range. Furthermore, we observed a more pronounced distance between the median of the box plot and the zero residual line at low redshifts compared to high redshifts. 
This observation implies larger deviations from zero residuals on average within this range, which may further complicate accurate predictions in lower redshift regions.

\begin{table*}
    \small
    \centering
    \begin{threeparttable}
    \caption{Comparison of CatBoost model performance on physical parameters with and without observational errors in input features, using the same dataset.}
    \label{tab:uncertainty}
    \begin{tabular}{lcccccccccc}
    \toprule
    \multirow{2}{*}{\textbf{Parameter}} & \multicolumn{4}{c}{\textbf{Without Errors}} & \multicolumn{6}{c}{\textbf{With Errors}} \\
    \cmidrule(lr){2-6} \cmidrule(lr){7-11}
     & \textbf{RMSE} & \textbf{NRMSE} & \textbf{MAE} & \textbf{Bias} & $\boldsymbol{\eta}$ & \textbf{RMSE} & \textbf{NRMSE} & \textbf{MAE} & \textbf{Bias} & $\boldsymbol{\eta}$ \\
    \midrule
    \multirow{1}{*}{SFR} & 0.353 & 0.040 & 0.263 & -0.010 & 0.011  & 0.336 & 0.038 & 0.250 & 0.001 & 0.011 \\    
    \multirow{1}{*}{SM} & 0.216 & 0.035 & 0.157 & 0.0003 & 0.011 & 0.206 & 0.033 & 0.148 & 0.001 & 0.012 \\      
    \multirow{1}{*}{metallicity} & 0.104 & 0.068 & 0.079 & -0.003 & 0.010 & 0.097 & 0.064 & 0.073 & -0.0002 & 0.011 \\    
    \bottomrule
    \end{tabular}
     \begin{tablenotes}
    \item \scriptsize \textbf{Note:} The SFR is measured in units of $ \rm{log(SFR[M_{\odot}yr^{-1}])} $, SM is also measured in units of 
    $ \rm{log(M^{\star}[M_{\odot}])} $, and metallicity is represented as $\rm{12+log(O/H)}$.
    \end{tablenotes}
    \end{threeparttable}
\end{table*}
\subsubsection{Enhanced predictive analysis: Exploring CatBoost with error considerations}\label{sec:error}
Our machine learning models are trained using observational data characterized via spectroscopy, a valid approach common in such studies. However, it is crucial to acknowledge that the accuracy of our presented models is inherently tied to the quality of the training data. 
Therefore, we have incorporated measurement uncertainties from the training data during the training process. This ensures that our models not only learn from the nominal values but also consider the potential variations and uncertainties in the data, contributing to a more robust and reliable predictive framework.

In the previous section, we concentrated on using magnitudes and colours as features in our predictive models. Now, we broaden our approach to consider the observational errors in our modelling process.
We opted for CatBoost for this experiment because it showed a better performance than WDNN.
Our main goal here is to thoroughly evaluate how magnitude uncertainties affect the predictive accuracy of our model.

To conduct this experiment, we removed entries with null errors and those where the errors exceeded 1. This reduces the available training data to 364~424 data points for SFR and SM prediction, and 121~060 data points for metallicity prediction.
We reevaluated our pre-trained CatBoost models from section \ref{sec:galaxies} using this updated dataset that now includes observational errors.
In contrast, for the models with errors, we developed new ML models trained specifically on this refined dataset. 
Following a similar approach to the previous section, 
we conducted hyperparameter tuning on CatBoost and 
then used 10-fold cross-validation exclusively on 
the training data. Through this process, we identified 
that a depth parameter of 10 resulted in optimal performance.

In Table \ref{tab:uncertainty}, we present the outcomes of this analysis and compare the performance of the CatBoost model with and without errors on the same dataset.
Our results show incorporating observational errors into the models yields enhanced performance compared to models that overlook such uncertainties.
Specifically, the RMSE for SFR decreases to 0.336~dex when errors are considered, compared to 0.353~dex without error considerations. Similarly, when examining the performance metrics with and without errors, we observe a reduction in RMSE for SM prediction, dropping from 0.216~dex to 0.206~dex. Likewise, for metallicity prediction, the RMSE decreases from 0.104~dex to 0.097~dex.
\begin{figure*}
  \centering
\subfloat{\includegraphics[width=0.25\linewidth]{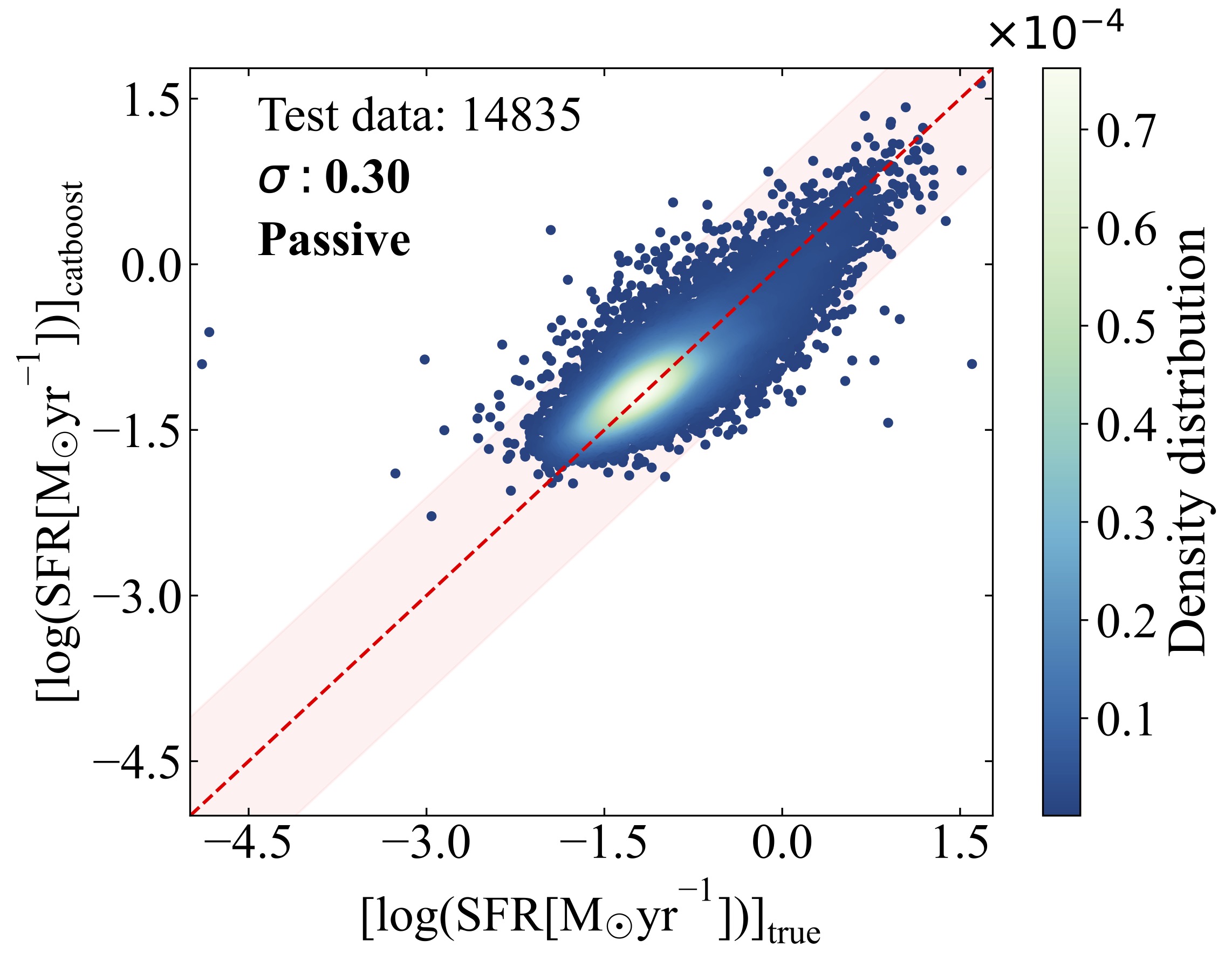}}\hfill 
  \subfloat{\includegraphics[width=0.25\linewidth]{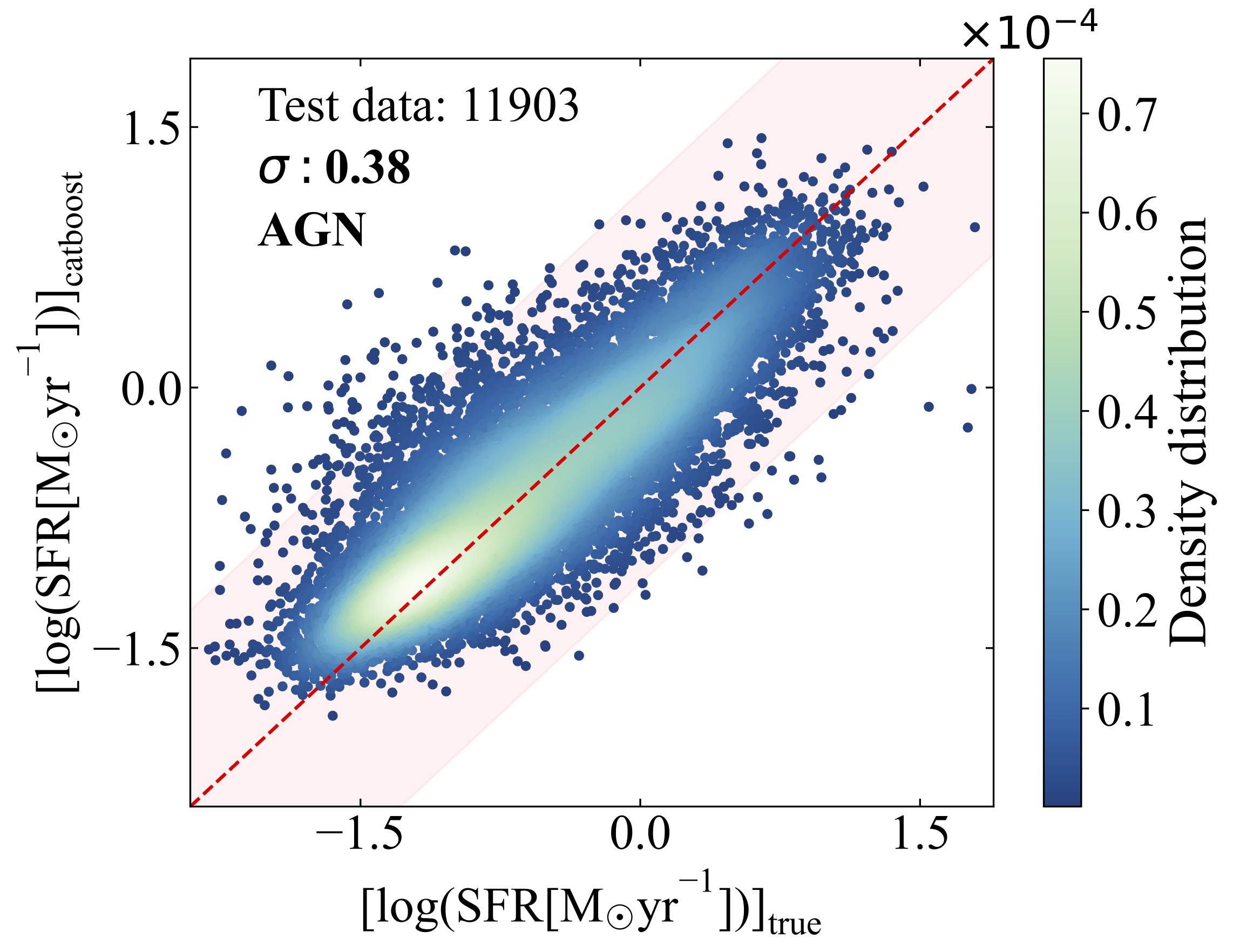}}\hfill
  \subfloat{\includegraphics[width=0.25\linewidth]{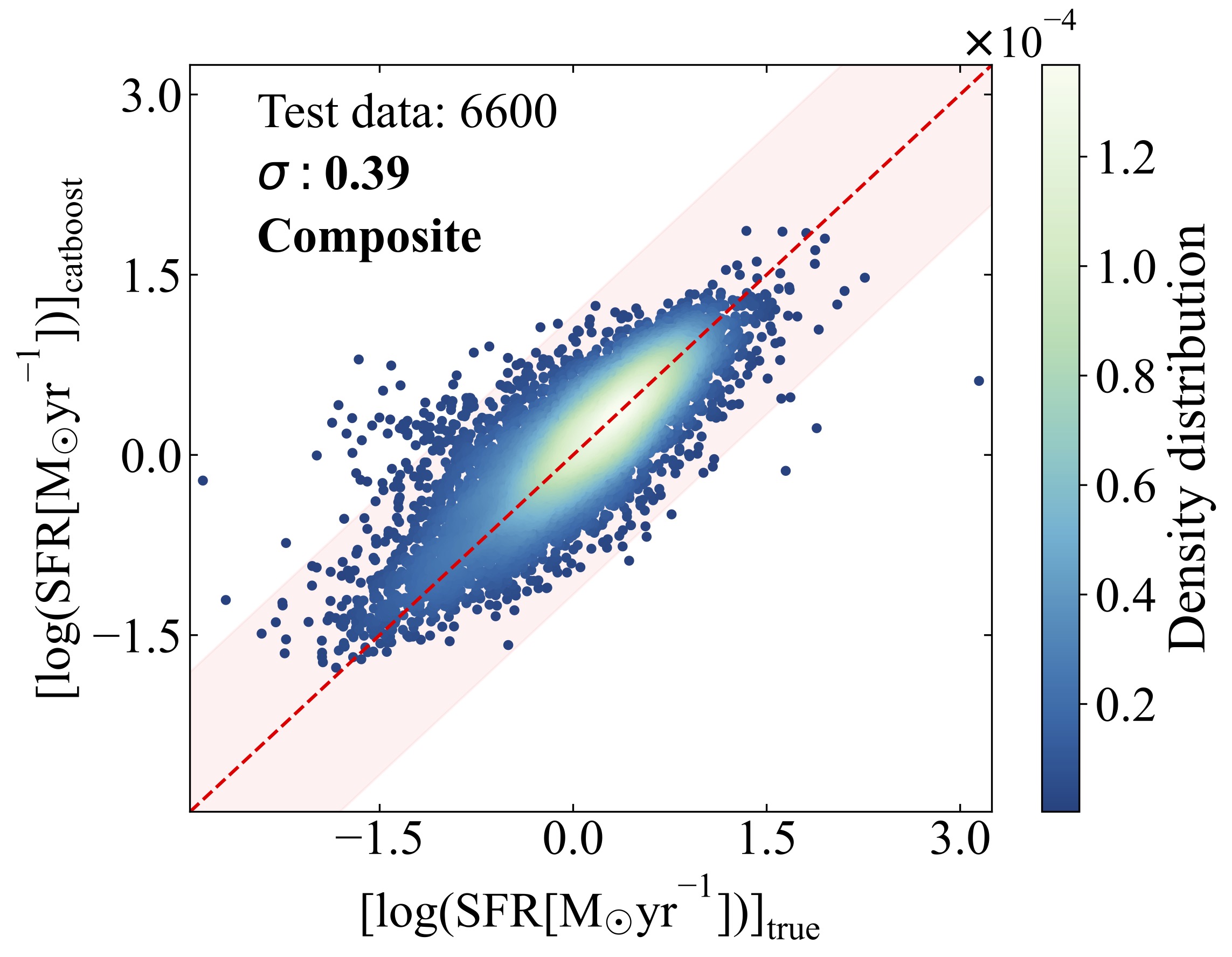}}\hfill
  \subfloat{\includegraphics[width=0.25\linewidth]{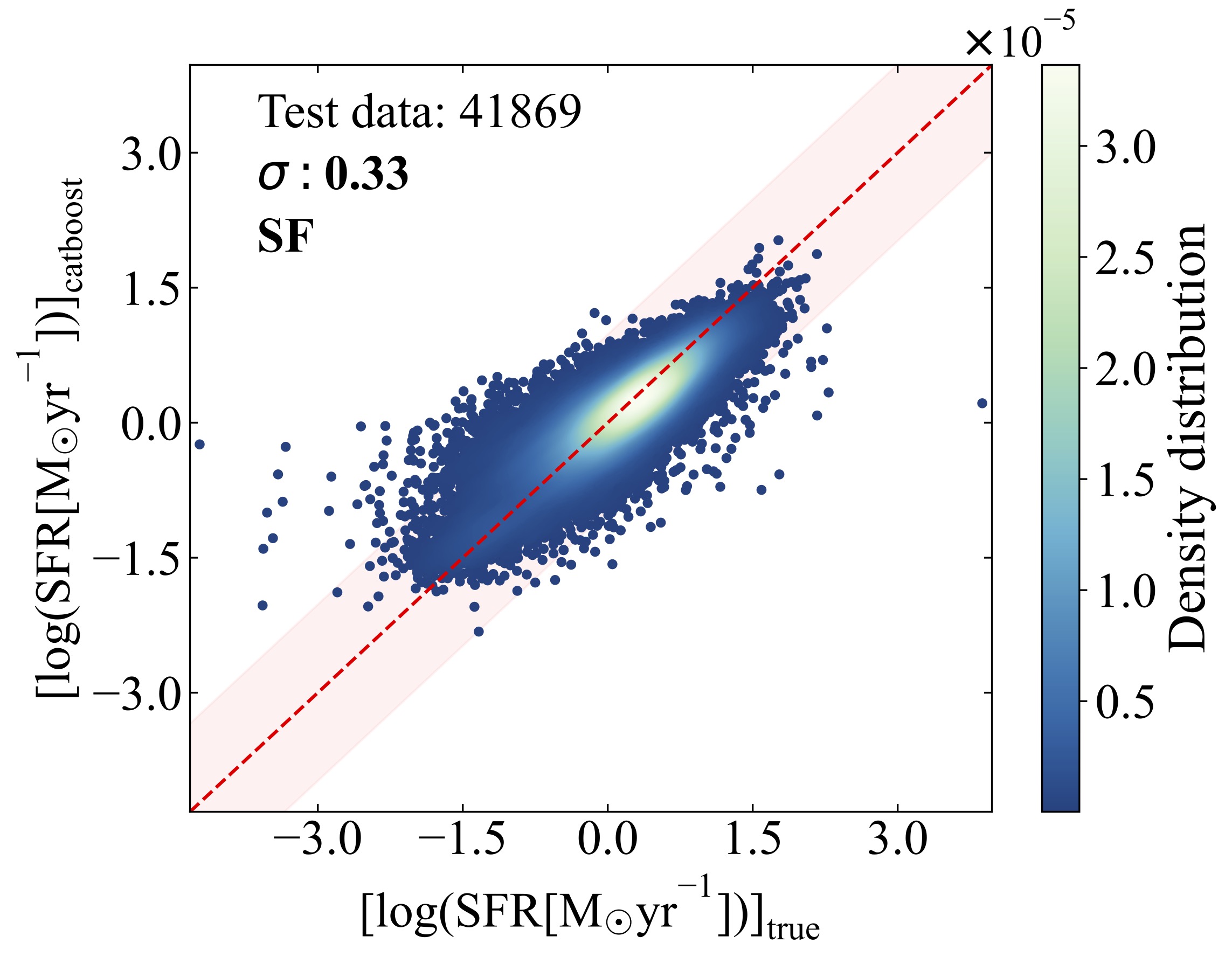}}\hfill
\subfloat{\includegraphics[width=0.25\linewidth]{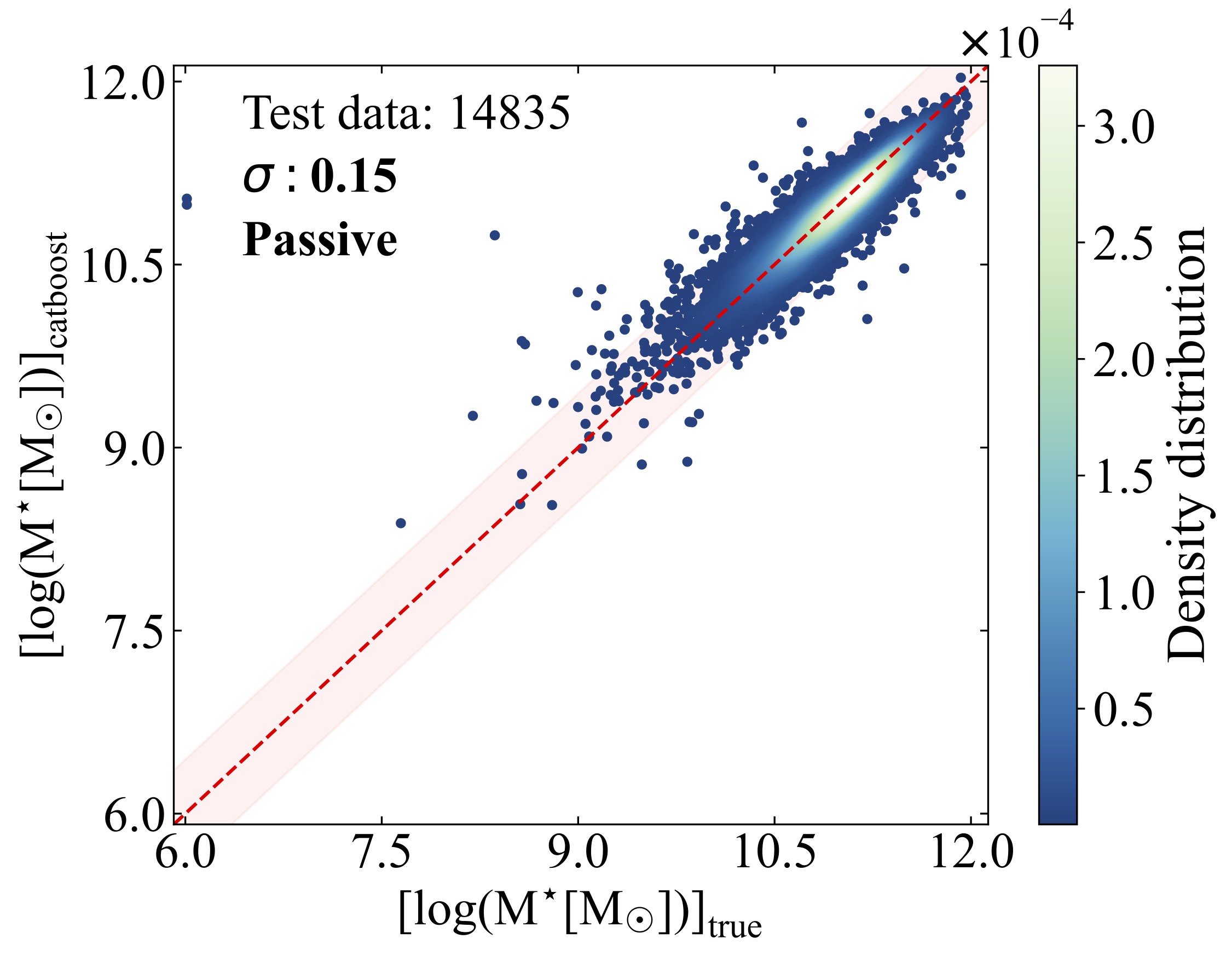}}\hfill 
  \subfloat{\includegraphics[width=0.25\linewidth]{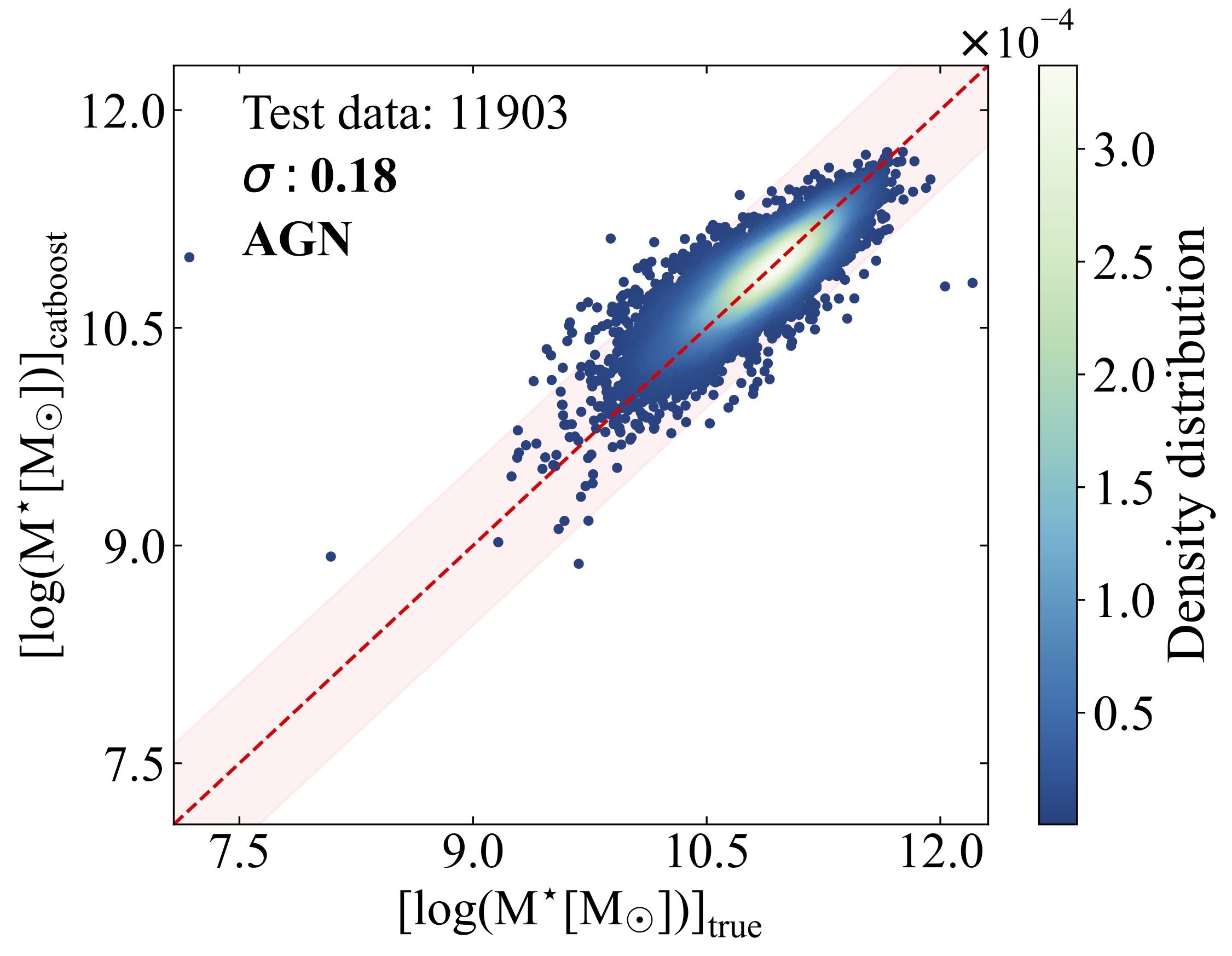}}\hfill
 \subfloat{\includegraphics[width=0.25\linewidth]{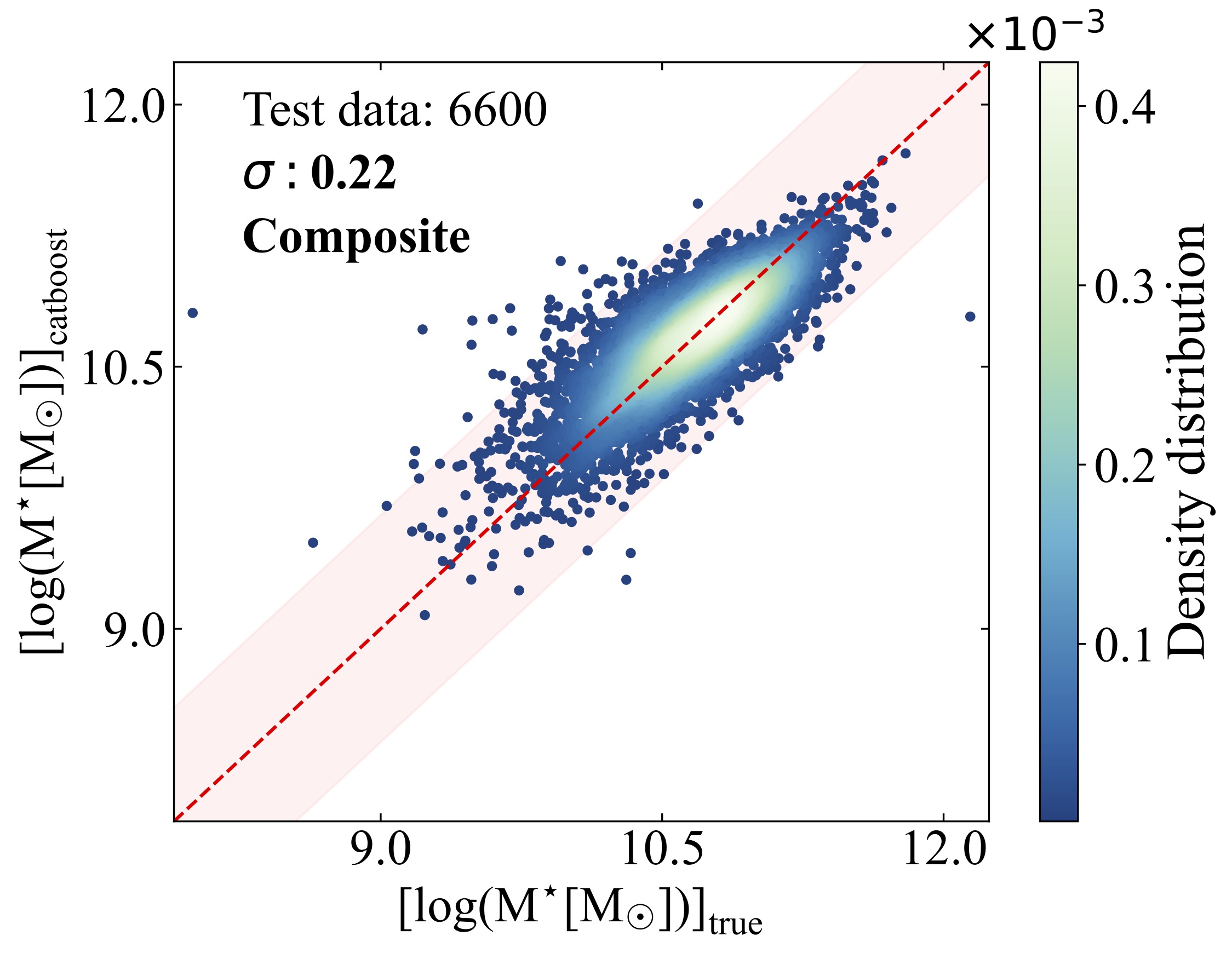}}\hfill
  \subfloat{\includegraphics[width=0.25\linewidth]{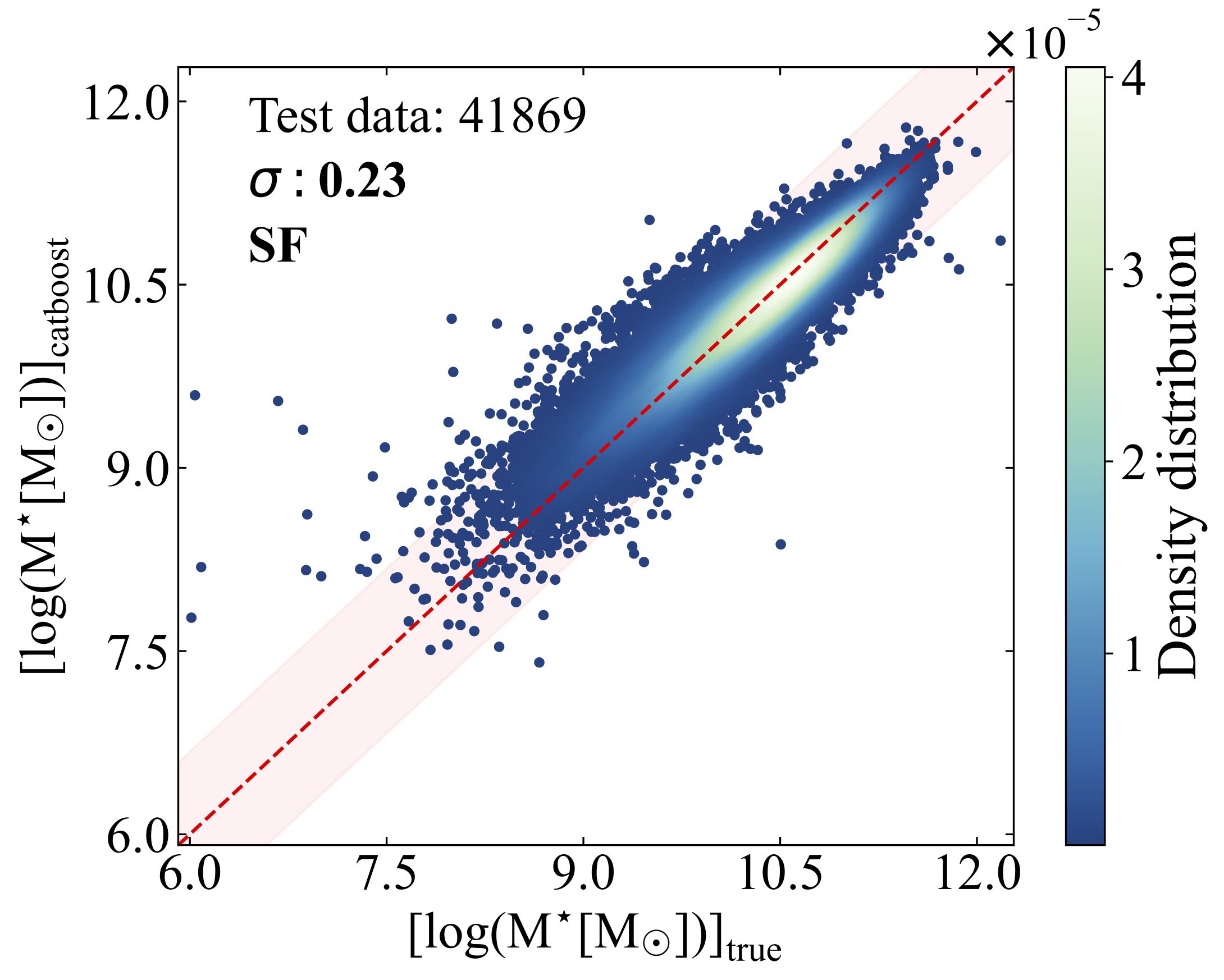}}\hfill
  \caption{Results of CatBoost in predicting SFR and SM for different kinds of galaxies.}
  \label{fig:bpt_cat}
\end{figure*}
\begin{table*}[h]
    \centering
    \begin{threeparttable}
    \caption{The performance of CatBoost model for different kinds of galaxies.}\label{tab:bpt}
    \begin{tabular}{ccccccc}
        \toprule
        \textbf{Class} & \textbf{Parameter} & \textbf{RMSE} & \textbf{NRMSE} & \textbf{MAE} & \textbf{Bias} & $\boldsymbol{\eta}$ \\
        \midrule
        \multirow{2}{*}{Passive} & SFR & 0.297 & 0.045 & 0.220  & 0.017 & 0.010 \\
        & SM & 0.149 & 0.025 & 0.099 & -0.018 & 0.009 \\
        \midrule
        \multirow{2}{*}{AGN} & SFR & 0.379 & 0.092 & 0.287 & 0.041 & 0.011 \\
        & SM & 0.184 & 0.037 & 0.134 & 0.012 & 0.012\\
        \midrule
        \multirow{2}{*}{Composite} & SFR & 0.391 & 0.065 & 0.288 & 0.049 & 0.013 \\
        & SM & 0.217 & 0.052 & 0.162 & 0.009 & 0.010 \\
        \midrule
        \multirow{2}{*}{Star-forming} & SFR & 0.327 & 0.040 & 0.244 & -0.023 & 0.011\\
        & SM & 0.225 & 0.036 & 0.167 & 0.003 & 0.011\\
        \bottomrule
    \end{tabular}
    \begin{tablenotes}
    \item \scriptsize \textbf{Note:} The SFR is measured in units of $ \rm{log(SFR[M_{\odot}yr^{-1}])} $, SM is also measured in units of $ \rm{log(M^{\star}[M_{\odot}])} $.
    \end{tablenotes}
    \end{threeparttable}
\end{table*}
\subsubsection{Performance of CatBoost for different kinds of galaxies}\label{sec:bpt}
In the preceding subsections, we conducted predictions without regard to galaxy type. 
This approach facilitated a comprehensive assessment of the predictive capabilities of 
our ML models and allowed us to evaluate their generalizability across a diverse array of galaxies.

To characterize the performance of our trained models more in detail,
we carried out focussed analyses for each specific group of galaxies in the current section.
Given the better performance of CatBoost compared to the WDNN,
we exclusively employed the CatBoost model, which accounts for observational errors, for this analysis.
Thus, we evaluated the predictive performance of the CatBoost algorithm based on 
the BPT classifications outlined in the SDSS MPA-JHU catalogue. Our test dataset 
was categorized into four groups: passive (14~835 data points), AGN (11~903 data points), 
composite (6~600 data points), and star-forming (41~869 data points). As the 
BPT classifications for galaxies with available metallicities were absent in 
the SDSS MPA-JHU catalogue, we did not analyse metallicity in this subsection.

We display the results of this analysis in Fig. \ref{fig:bpt_cat}, highlighting the difference between the predicted
values of SFR (upper row subplots) and SM (bottom row subplots) generated by CatBoost and the true values from the SDSS MPA-JHU catalogue. The comparisons are conducted across various BPT classes, organised from left to right as passive, AGNs, composites, and star-forming galaxies.
The robust consistency between the true values and the predicted values is evident across various galaxy categories.

Table \ref{tab:bpt} provides a detailed overview of the evaluation metrics, encompassing RMSE, NRMSE, MAE, bias, and $ \eta $ specifically across galaxy types. 
We observe an overall good performance across all types of galaxies, but a closer analysis of the results reveals significant variations in predictive accuracy, especially in the transition from passive to more active galaxies.
This transition significantly impacts both SFR and SM predictions.

The NRMSE values for SFR prediction indicate that star-forming galaxies exhibit the lowest NRMSE at 0.040, passive galaxies at 0.045, composite galaxies at 0.065, and AGNs at 0.092. Likewise, for SM prediction, passive galaxies demonstrate the most accurate prediction with an NRMSE of 0.025, followed by star-forming galaxies at 0.036, AGN galaxies at 0.037, and composite galaxies at 0.052. The higher evaluation metrics in AGNs, composites, and star-forming galaxies suggest increased prediction errors for both SFR and SM compared to passive galaxies.

\begin{table*}[h]
\centering
\caption{Model performance comparison with related studies.}\label{tab:comparison}
\resizebox{\textwidth}{!}{%
\begin{threeparttable}
\begin{tabular}{llccccccl}
\toprule
\textbf{Model} & \textbf{Data} & \textbf{RMSE$_{\text{SFR}}$} & \textbf{RMSE$_{\text{SM}}$} & \textbf{RMSE$_{\text{metallicity}}$} & \textbf{RMSE$_{\text{sSFR}}$} & \textbf{Reference} \\ \midrule
CatBoost & SDSS\&WISE & 0.336 & 0.206 & 0.097 & 0.248 & \textbf{This work} \\ 
Random Forest & SDSS\&WISE & 0.38 & 0.16 & - & -  & B19 \\ 
Random Forest & SDSS & - & - & - & 0.252 & DV19 \\ 
DL & GAMA & 0.1658 & 0.0582 & - & - & S20 \\ \bottomrule
\end{tabular}%
\begin{tablenotes}
\item \scriptsize \textbf{Note:} The SFR is measured in units of $ \rm{log(SFR[M_{\odot}yr^{-1}])} $, SM is also measured in units of $ \rm{log(M^{\star}[M_{\odot}])} $, and metallicity is represented as $\rm{12+log(O/H)}$.
\item \scriptsize \textbf{References:} (1) Bonjean et al. (2019) [B19]; (2) Delli Veneri et al. (2019) [DV19]; (3) Surana et al. (2020) [S20].
\end{tablenotes}
\end{threeparttable}}
\end{table*}
\subsubsection{Comparison to other works}

At the outset of the comparison with other works, it is crucial to acknowledge the inherent variations among studies arising from the utilization of different data and input patterns. Table \ref{tab:comparison} only provides a simple comparative analysis of methodologies employed in the automated prediction of galaxy properties, as explored by \cite{Bonjean2019, DelliVeneri2019, Surana2020}. 
Additionally, we include our best-performing model, the CatBoost model that accounts for observational errors, in this table for comparison.
\cite{Bonjean2019} conducted a comprehensive study using the SDSS catalogue, and cross-matched it with AllWISE catalogue. Their input features comprised IR magnitudes and spectroscopic redshifts, and they employed a random forest algorithm trained on SFR and SM data extracted from the SDSS MPA-JHU catalogue. 
Since the RMSE and \(\sigma\) values in our results are identical (because the mean of $\Delta X_{i}$ in eq. \ref{eq:sigma} effectively vanishes), we specify that the \(\sigma\) reported by Bonjean has been incorporated into the RMSE column of this table.
Their reported errors were $\rm{\sigma_{SFR}} = 0.38$~dex and $ \rm{\sigma_{SM}} = 0.16$~dex, whereas our model yielded 0.336~dex and 0.206~dex, respectively.
\cite{DelliVeneri2019} used 55 photometric features (magnitudes, colours, and $photoz$). Their method involved estimating specific SFR ($ \rm{sSFR = SFR/M_{\star}} $) using techniques like parameter space optimization and employing a supervised ML model, such as a random forest. Their reported error for sSFR was $\rm{RMSE_{sSFR}=0.252}$~dex. Compared to these two works, we applied 8 optical and infrared magnitudes and colours ($r, u-g, g-r, r-i, i-z, z-W1, W1-W2, W2-W3$), and only based on photometric information without spectroscopic redshifts, $\rm{RMSE_{sSFR}}$ arrived at 0.248~dex.

\cite{Surana2020} utilized data from the Galaxy And Mass Assembly (GAMA) survey, which incorporates a wide range of input features, such as from Far-UV, near-UV, optical, near-IR, mid-IR bands, and spectroscopic redshifts. They employed DL techniques to predict three properties of galaxies.
Two of our galaxy properties align with two of theirs, namely, SFR and SM. Comparing the results of predicting these two parameters, the reported errors for their predictions were that $\rm{RMSE_{SFR}} = 0.1658$~dex and $\rm{RMSE_{SM}} = 0.0582$~dex. In comparison, our model produced errors of 0.336~dex and 0.206~dex for the same properties, respectively. 
This suggests that, as indicated by \cite{Surana2020}, having additional information from multiple bands, specifically from UV, optical and IR, enhances the accuracy of estimating physical parameters.

\section{Summary and conclusions}\label{summary}
In this research, we explored the application of ML models, specifically utilizing CatBoost and WDNN, to predict essential galaxy properties — SFR, SM, and metallicity. Our approaches were trained using optical and IR photometric data from SDSS DR18 and AllWISE, with the SDSS MPA-JHU DR8 values serving as a target for training. 
In addition, we incorporated observational errors into our models to offer a comprehensive assessment of the accuracy achievable with the trained ML methods.

In our methodology, we adopted a dual-model strategy for both of the ML techniques under investigation. The first model was configured to accommodate two outputs, specifically the prediction of SFR and SM. 
For the second model, our focus was on training it with a singular output, specifically predicting metallicity. Moreover, we introduced innovative DL architectures, identifying two effective WDNN structures that achieved competitive results compared with prior studies.

Notably, during this analysis, we found that CatBoost outperformed WDNN. To further explore the impact of observational uncertainties on our predictions, we conducted an additional experiment incorporating observational errors alongside magnitudes and colours, using CatBoost exclusively. The results indicated significantly improved performance when observational errors were considered, particularly evident across all three galaxy parameters compared to when these errors were not factored in. This finding suggests the importance of addressing observational data uncertainties for achieving improved predictive accuracy.

Subsequently, we conducted a thorough analysis of various galaxy types categorized by the BPT classifications in the same catalogue. The galaxies were divided into passive and active groups, encompassing AGN, composite, and star-forming categories. 
In this test our CatBoost model that accounts for observational errors demonstrates its robustness, providing a reliable performance across all considered galaxy types. We do, however, observe a decrease in predictive performance for the more active galaxies.

In conclusion, our study underscores the adaptability of both the CatBoost and WDNN models in predicting galaxy properties using optical and IR photometric data. The expanding volume of astronomical data from larger surveys has created a growing demand for accurate and timely predictions, making ML a particularly suitable solution. Our models demonstrate the capability to provide predictions with a high degree of certainty, offering promising avenues for further exploration in ongoing and upcoming photometric surveys such as James Webb Space Telescope (JWST), LSST, and Euclid. Opportunities for improvement lie in integrating information derived from broad surveys with comprehensive observations.

\section*{Acknowledgements}
We specially thank the reviewer for his invaluable feedback and insightful comments to help us improve our paper. This paper is funded by the National Natural Science Foundation of China under grant No. 12273076 (Y.Z.) and the science research grants from the China Manned Space Project with Nos. CMS-CSST-2021-A04 and CMS-CSST-2021-A06.

We acknowledge the use of SDSS  and WISE databases.
Funding for the Sloan Digital Sky Survey IV has been provided by the Alfred P. Sloan Foundation, the U.S. Department of Energy Office of Science, and the Participating Institutions. SDSS-IV acknowledges
support and resources from the Centre for High-Performance Computing at
the University of Utah. The SDSS website is www.sdss.org.
Funding for the SDSS has been provided by the Alfred P.
Sloan Foundation, the Participating Institutions, the National
Science Foundation, the U.S. Department of Energy, the National Aeronautics and Space Administration, the Japanese
Monbukagakusho, the Max Planck Society, and the Higher
Education Funding Council for England. The SDSS Web
Site is http://www.sdss.org/.

The SDSS is managed by the Astrophysical Research Consortium for the Participating Institutions. The Participating
Institutions are the American Museum of Natural History,
Astrophysical Institute Potsdam, University of Basel, University of Cambridge, Case Western Reserve University, University of Chicago, Drexel University, Fermilab, the Institute for
Advanced Study, the Japan Participation Group, Johns Hopkins University, the Joint Institute for Nuclear Astrophysics,
the Kavli Institute for Particle Astrophysics and Cosmology, the Korean Scientist Group, the Chinese Academy
of Sciences (LAMOST), Los Alamos National Laboratory,
the Max-Planck-Institute for Astronomy (MPIA), the MaxPlanck-Institute for Astrophysics (MPA), New Mexico State
University, Ohio State University, University of Pittsburgh,
University of Portsmouth, Princeton University, the United
States Naval Observatory, and the University of Washington.

This publication makes use of data products from the Wide-field Infrared Survey Explorer, which is a joint project of the University of California, Los Angeles, and the Jet Propulsion Laboratory/California Institute of Technology, funded by the National Aeronautics and Space Administration.

%

\begin{thebibliography}{}
\bibitem[\protect\citeauthoryear{Abraham et~al.}{2012}]{Abraham2012}
Abraham, S., Philip, N. S., Kembhavi, A., Wadadekar, Y. G., \& Sinha, R. 2012, \mnras , 419, 80
\bibitem[\protect\citeauthoryear{Abraham et~al.}{2018}]{Abraham2018}
Abraham, S., Aniyan, A. K., Kembhavi, A. K., Philip, N. S., \& Vaghmare, K. 2018, \mnras , 477, 894
\bibitem[\protect\citeauthoryear{Allen et~al.}{2019}]{Allen2019}
Allen, G., Andreoni, I., Bachelet, E., et al. 2019, arXiv preprint arXiv:1902.00522
\bibitem[\protect\citeauthoryear{Baldry et~al.}{2008}]{Baldry2008}
Baldry, I. K., Glazebrook, K., \& Driver, S. P. 2008, \mnras, 388, 945
\bibitem[\protect\citeauthoryear{Baldwin et~al.}{1981}]{Baldwin1981}
Baldwin, J. A., Phillips, M. M., \& Terlevich, R. 1981, \pasp, 93, 5
\bibitem[\protect\citeauthoryear{Ball \& Brunner}{2010}]{Ball2010}
Ball, N. M., \& Brunner, R. J. 2010, International Journal of Modern Physics D, 19, 1049
\bibitem[\protect\citeauthoryear{Ball et~al.}{2006}]{Ball2006}
Ball, N. M., Brunner, R. J., Myers, A. D., \& Tcheng, D. 2006, \apj , 650, 497
\bibitem[\protect\citeauthoryear{Balogh et~al.}{1999}]{Balogh1999}
Balogh, M. L., Morris, S. L., Yee, H. K. C., Carlberg, R. G., \& Ellingson, E. 1999, \apj, 527, 54
\bibitem[\protect\citeauthoryear{Barchi et~al.}{2020}]{Barchi2020}
Barchi, P. H., de Carvalho, R., Rosa, R., et~al. 2020, Astron. Comput. , 30, 100334
\bibitem[\protect\citeauthoryear{Bishop}{2006}]{Bishop2006}
Bishop, C. M. 2006, Pattern Recognition and Machine Learning (Springer)
\bibitem[\protect\citeauthoryear{Bisigello et~al.}{2016}]{Bisigello2016}
Bisigello L., Caputi, K. I., Colina, L., et~al. 2016, \apjs, 227, 19 
\bibitem[\protect\citeauthoryear{Bisigello et~al.}{2017}]{Bisigello2017}
Bisigello L., Caputi, K. I., Colina, L.,  et~al. 2017, \apjs, 231, 3
\bibitem[\protect\citeauthoryear{Bonjean et~al.}{2019}]{Bonjean2019}
Bonjean, V., Aghanim, N., Salom\'{e}, P., et~al. 2019, \aa, 622, A137
\bibitem[\protect\citeauthoryear{Brescia et~al.}{2014}]{Brescia2014}
Brescia, M., Cavuoti, S., Longo, G., \& De Stefano, V., 2014, \aap, 568, A126
\bibitem[\protect\citeauthoryear{Brinchmann et~al.}{2004}]{Brinchmann2004}
Brinchmann, J., Charlot, S., White, S. D. M., et~al. 2004, \mnras, 351, 1151
\bibitem[\protect\citeauthoryear{Bruzual \& Charlot}{2003}]{Bruzual2003}
Bruzual, G., \& Charlot, S. 2003, \mnras, 344, 1000
\bibitem[\protect\citeauthoryear{Bruzual}{1983}]{Bruzual1983}
Bruzual, A. G. 1983, \apj, 273, 105
\bibitem[\protect\citeauthoryear{Calzetti et~al.}{1994}]{Calzetti1994}
Calzetti, D., Kinney, A. L., \& Storchi-Bergmann, T. 1994, \apj, 429, 582
\bibitem[\protect\citeauthoryear{Cheng et~al.}{2016}]{Cheng2016}
Cheng, H.-T., et al., “Wide \& Deep Learning for Recommender Systems,” Proceedings of the First Workshop on Deep Learning for Recommender Systems (2016): 7–10
\bibitem[\protect\citeauthoryear{Cheng et~al.}{2020}]{Cheng2020}
Cheng, T.-Y., Li, N., Conselice, C. J., et~al. 2020, \mnras , 494, 3750
\bibitem[\protect\citeauthoryear{Chollet et~al.}{2015}]{Chollet2015}
Chollet, F., et~al. 2015, Keras, https://keras.io
\bibitem[\protect\citeauthoryear{Ciesla et~al.}{2017}]{Ciesla2017}
Ciesla, L., Elbaz, D., \& Fensch, J. 2017, \aap, 608, A41
\bibitem[\protect\citeauthoryear{Clarke et~al.}{2020}]{Clarke2020}
Clarke, A. O., Scaife, A. M. M., Greenhalgh, R., \& Griguta, V., 2020, \aap, 639, A84
\bibitem[\protect\citeauthoryear{Conroy}{2013}]{Conroy2013}
Conroy, C. 2013, \araa, 51, 393
\bibitem[\protect\citeauthoryear{Cunha \& Humphrey}{2022}]{Cunha2022}
Cunha, P. A. C., \& Humphrey, A., 2022, \aap, 666, A87
\bibitem[\protect\citeauthoryear{D'Isanto \& Polsterer}{2018}]{DIsanto2018}
D'Isanto, A., \& Polsterer, K. L. 2018, \aap, 609, A111
\bibitem[\protect\citeauthoryear{Delli Veneri et~al.}{2019}]{DelliVeneri2019}
Delli Veneri, M., Cavuoti, S., Brescia, M., Longo, G., \& Riccio, G. 2019, \mnras, 486, 1377
\bibitem[\protect\citeauthoryear{Dieleman et~al.}{2015}]{Dieleman2015}
Dieleman, S., Willett, K. W., \& Dambre, J. 2015, \mnras , 450, 1441
\bibitem[\protect\citeauthoryear{ Dom\'{i}nguez S\'{a}nchez et~al.}{2018}]{Dom2018}
Dom\'{i}nguez S\'{a}nchez, H., Huertas-Company, M., Bernardi, M., Tuccillo, D., \& Fischer, J. L. 2018, \mnras , 476, 3661
\bibitem[\protect\citeauthoryear{Dorogush et~al.}{2018}]{Dorogush2018}
Dorogush, A. V., Ershov, V., \& Yandex, A. G. 2018, preprint (arXiv:1810.11363)
\bibitem[\protect\citeauthoryear{Dressler et~al.}{1987}]{Dressler1987}
Dressler A., Lynden-Bell, D., Burstein, D., et~al. 1987, \apj , 313, 42  
\bibitem[\protect\citeauthoryear{Faber}{1973}]{Faber1973}
Faber, S. M. 1973, \apj, 179, 731
\bibitem[\protect\citeauthoryear{Fogarty et~al.}{2017}]{Fogarty2017}
Fogarty, K., Postman, M., Larson, R., Donahue, M., \& Moustakas, J., 2017, \apj, 846, 103
\bibitem[\protect\citeauthoryear{Friedman}{2001}]{Friedman2001}
Friedman, J. H. 2001, Ann. Stat., 29, 1189
\bibitem[\protect\citeauthoryear{Gallazzi et~al.}{2005}]{Gallazzi2005}
Gallazzi, A., Charlot, S., Brinchmann, J., White, S. D. M., \& Tremonti, C. A. 2005, \mnras, 362, 41
\bibitem[\protect\citeauthoryear{Garnett}{2002}]{Garnett2002}
Garnett, D. R. 2002, \apj, 581, 1019
\bibitem[\protect\citeauthoryear{George \& Huerta}{2018}]{George2018}
George, D., \& Huerta, E. A. 2018, Phys. Lett. B, 778, 64
\bibitem[\protect\citeauthoryear{Goodfellow et~al.}{2016}]{Goodfellow2016}
Goodfellow, I., Bengio, Y., \& Courville, A. 2016, Deep Learning (MIT Press)
\bibitem[\protect\citeauthoryear{Hoyle}{2016}]{Hoyle2016}
Hoyle, B. 2016, Astronomy and Computing, 16, 34
\bibitem[\protect\citeauthoryear{Ivezi\'{c} et~al.}{2019}]{Ivezic2019}
Ivezi\'{c}, \v{Z}. et~al., 2019, \apj , 873, 111
\bibitem[\protect\citeauthoryear{Janowiecki et~al.}{2017}]{Janowiecki2017}
Janowiecki, S., Catinella, B., Cortese, L., et~al. 2017, \mnras, 466, 4795
\bibitem[\protect\citeauthoryear{Jones et~al.}{2016}]{Jones2016}
Jones, M. L., Hickox, R. C., Black, C. S., et~al. 2016, \apj, 826, 12
\bibitem[\protect\citeauthoryear{Kauffmann et~al.}{2003}]{Kauffmann2003}
Kauffmann, G., Heckman, T. M., White, S. D. M., et~al. 2003, \mnras, 341, 33
\bibitem[\protect\citeauthoryear{Kennicutt}{1998}]{Kennicutt1998}
Kennicutt, Jr. R. C.  1998, \apj, 498, 541
\bibitem[\protect\citeauthoryear{Kennicutt \& Evans}{2012}]{Kennicutt2012}
Kennicutt, R. C., \& Evans, N. J. 2012, \araa, 50, 531
\bibitem[\protect\citeauthoryear{Kim \& Brunner}{2017}]{Kim2017}
Kim, E. J., \& Brunner, R. J. 2017, \mnras, 464, 4463
\bibitem[\protect\citeauthoryear{Kohavi}{1995}]{Kohavi1995}
Kohavi, R. A Study of Cross-validation and Bootstrap for Accuracy Estimation and Model Selection. In Proceedings of the 14th International Joint Conference on Artificial Intelligence - Volume 2, IJCAI'95, 1995.
\bibitem[\protect\citeauthoryear{Kormendy}{1977}]{Kormendy1977}
Kormendy, J. 1977, \apj, 218, 333
\bibitem[\protect\citeauthoryear{Krakowski et~al.}{2016}]{Krakowski2016}
Krakowski, T., Małek, K., Bilicki, M., et~al. 2016, \aap, 596, A39
\bibitem[\protect\citeauthoryear{Kravtsov et~al.}{2018}]{Kravtsov2018}
Kravtsov, A. V., Vikhlinin, A. A., \& Meshcheryakov, A. V. 2018, Astron. Lett., 44, 8
\bibitem[\protect\citeauthoryear{Kroupa}{2001}]{Kroupa2001}
Kroupa, P. 2001, \mnras, 322, 231
\bibitem[\protect\citeauthoryear{Lagache et~al.}{2005}]{Lagache2005}
Lagache, G., Puget, J.-L., \& Dole, H. 2005, \araa, 43, 727
\bibitem[\protect\citeauthoryear{Laigle et~al.}{2019}]{Laigle2019}
Laigle, C., Davidzon, I., Ilbert, O., et~al. 2019, \mnras, 486, 5104
\bibitem[\protect\citeauthoryear{Lara-Lopez et~al.}{2010}]{LaraLopez2010}
Lara-Lopez, M. A., Cepa, J., Bongiovanni, A., et~al. 2010, \aap, 521, L53
\bibitem[\protect\citeauthoryear{Laureĳs et~al.}{2011}]{LaurIJs2011}
Laureĳs, R., et~al., 2011, arXiv:1110.3193
\bibitem[\protect\citeauthoryear{Leger \& Puget}{1984}]{Leger1984}
Leger, A., \& Puget, J. L. 1984, \aap, 137, L5
\bibitem[\protect\citeauthoryear{Lequeux et~al.}{1979}]{Lequeux1979}
Lequeux, J., Peimbert, M., Rayo, J. F., Serrano, A., \& Torres-Peimbert S. 1979, \aap, 80, 155
\bibitem[\protect\citeauthoryear{Li et~al.}{2023}]{Li2023}
Li, C., Zhang, Y., Chenzhou, C., et~al. 2023, \mnras, 518, 513
\bibitem[\protect\citeauthoryear{Mannucci et~al.}{2010}]{Mannucci2010}
Mannucci, F., Cresci, G., Maiolino, R., Marconi, A., \& Gnerucci, A. 2010, \mnras, 408, 2115
\bibitem[\protect\citeauthoryear{Masters et~al.}{2015}]{Masters2015}
Masters, D., Capak, P., Stern, D., et~al. 2015, \apj, 813, 53
\bibitem[\protect\citeauthoryear{Masters et~al.}{2019}]{Masters2019}
Masters, D. C., Stern, D. K., Cohen, J. G., et~al. 2019, \apj, 877, 81
\bibitem[\protect\citeauthoryear{Mitchell et~al.}{2013}]{Mitchell2013}
Mitchell, P. D., Lacey, C. G., Baugh, C. M., \& Cole, S. 2013, \mnras,435,87
\bibitem[\protect\citeauthoryear{Mobasher et~al.}{2015}]{Mobasher2015}
Mobasher, B., Dahlen, T., Ferguson, H. C., et~al. 2015, \apj, 808, 101
\bibitem[\protect\citeauthoryear{Mucesh et~al.}{2021}]{Mucesh2021}
Mucesh S., Hartley, W. G., Palmese, A., et~al. 2021, \mnras , 502, 2770
\bibitem[\protect\citeauthoryear{Nair \& Vivek}{2022}]{Nair2022}
Nair, A., \& Vivek, M. 2022, \mnras , 511, 4946
\bibitem[\protect\citeauthoryear{Nakoneczny et~al.}{2021}]{Nakoneczny2021}
Nakoneczny, S. J., Bilicki, M., Pollo1, A., et~al., 2021, \aap, 649, A81
\bibitem[\protect\citeauthoryear{Pacifici et~al.}{2015}]{Pacifici2015}
Pacifici, C., da Cunha, E., Charlot, S., et~al. 2015, \mnras, 447, 786
\bibitem[\protect\citeauthoryear{Pearson et~al.}{2018}]{Pearson2018}
Pearson, W. J., Wang, L., Hurley, P. D., et~al., 2018, \aap, 615, A146
\bibitem[\protect\citeauthoryear{Philip et~al.}{2002}]{Philip2002}
Philip, N. S., Wadadekar, Y., Kembhavi, A., \& Joseph, K. B. 2002, \aap , 385, 1119
\bibitem[\protect\citeauthoryear{Razim et~al.}{2021}]{Razim2021}
Razim, O., Cavuoti, S., Brescia, M., et~al. 2021, \mnras , 507, 5034
\bibitem[\protect\citeauthoryear{Roberts \& Haynes}{1994}]{Roberts1994}
Roberts, M. S., \& Haynes, M. P. 1994, \araa, 32, 115
\bibitem[\protect\citeauthoryear{Salim et~al.}{2007}]{Salim2007}
Salim, S., Rich, R. M., Charlot, S., et~al. 2007, \apjs, 173, 267
\bibitem[\protect\citeauthoryear{Salvato et~al.}{2019}]{Salvato2019}
Salvato, M., Ilbert, O., \& Hoyle, B. 2019, Nat. Astron. 3, 212
\bibitem[\protect\citeauthoryear{Schindler et~al.}{2017}]{Schindler2017}
Schindler, J. T., Fan, X., McGreer, I. D., et~al. 2017, \apj, 851, 13
\bibitem[\protect\citeauthoryear{Siudek et~al.}{2018}]{Siudek2018}
Siudek, M., Małek, K., Pollo, A. et~al. 2018, \aap, 617, A70
\bibitem[\protect\citeauthoryear{Smith \& Hayward}{2015}]{Smith2015}
Smith, D. J. B., \& Hayward, C. C. 2015, \mnras, 453, 1597
\bibitem[\protect\citeauthoryear{Soumagnac et~al.}{2015}]{Soumagnac2015}
Soumagnac, M. T., Abdalla, F. B., Lahav, O., et~al. 2015, \mnras , 450, 666
\bibitem[\protect\citeauthoryear{Speagle et~al.}{2016}]{Speagle2016}
Speagle, J. S., Capak, P. L., Eisenstein, D. J., Masters, D. C., \& Steinhardt, C. L. 2016, \mnras, 461, 3432
\bibitem[\protect\citeauthoryear{Stensbo-Smidt et~al.}{2017}]{Stensbo-Smidt2017}
Stensbo-Smidt, K., Gieseke, F., Igel, C., Zirm, A., \& Steenstrup Pedersen, K., 2017, \mnras, 464, 2577
\bibitem[\protect\citeauthoryear{Su et~al.}{2013}]{Su2013}
Su, S., Kong, X., Li, J., \& Fang, G., 2013, \apj, 778, 10
\bibitem[\protect\citeauthoryear{Surana et~al.}{2020}]{Surana2020}
Surana, S., Wadadekar, Y., Bait, O., \& Bhosale, H. 2020, \mnras, 493, 4808
\bibitem[\protect\citeauthoryear{Tagliaferri et~al.}{2003}]{Tagliaferri2003}
Tagliaferri, R., Longo, G., Andreon, S., et~al. 2003, Lecture Notes in Computer Science, 2859, 226
\bibitem[\protect\citeauthoryear{Tremonti et~al.}{2004}]{Tremonti2004}
Tremonti, C. A., Heckman, T. M., Kauffmann, G., et~al. 2004, \apj, 613, 898 
\bibitem[\protect\citeauthoryear{Turner et~al.}{2019}]{Turner2019}
Turner, S., Kelvin, L. S., Baldry, I. K., et~al. 2019, \mnras, 482, 126
\bibitem[\protect\citeauthoryear{Vasconcellos et~al.}{2011}]{Vasconcellos2011}
Vasconcellos, E. C., de Carvalho, R. R., Gal, R. R., et~al. 2011, \aj , 141, 189
\bibitem[\protect\citeauthoryear{Walcher et~al.}{2011}]{Walcher2011}
Walcher, J., Groves, B., Budav\'{a}ri, T., \& Dale, D. 2011, \apss, 331, 1
\bibitem[\protect\citeauthoryear{Walmsley et~al.}{2020}]{Walmsley2020}
Walmsley, M., Smith, L., Lintott, C., et~al. 2020, \mnras, 491, 1554
\bibitem[\protect\citeauthoryear{Wijesinghe et~al.}{2007}]{Wijesinghe2012}
Wijesinghe, D. B., Hopkins, A. M., Brough, S., et~al. 2012, \mnras , 423, 3679
\bibitem[\protect\citeauthoryear{Wright et~al.}{2010}]{Wright2010}
Wright, E. L., Eisenhardt, P. R. M., Mainzer, A. K., et~al. 2010, \aj, 140, 1868
\bibitem[\protect\citeauthoryear{York et~al.}{2000}]{York2000}
York, D. G., Adelman, J., Anderson, Jr., J. E., et~al. 2000, \aj, 120, 1579
\bibitem[\protect\citeauthoryear{Zeraatgari et~al.}{2023}]{Zeraatgari2023}
Zeraatgari, F. Z., Hafezianzadeh, F., Zhang, et~al. 2024, \mnras, 527, 4677
\bibitem[\protect\citeauthoryear{Zhan}{2011}]{Zhan2011}
Zhan, H. 2011, SSPMA, 41, 1441
%
                                 %
\end{thebibliography}
%

\end{document}